\documentclass[journal]{acmart}
\usepackage{array}
\usepackage{comment}
\usepackage{pifont} 
\usepackage[utf8]{inputenc}
\usepackage{color,soul}

\newcommand{\cmark}{\ding{51}} 
\newcommand{\xmark}{\ding{55}} 
\AtBeginDocument{%
  }

\begin{document}

\title{Mining Q\&A Platforms for Empirical Evidence on Quantum Software Programming}
\author{Arif Ali Khan}
\email{arif.khan@oulu.fi}
\orcid{0000-0002-8479-1481}
\authornotemark[1]
\affiliation{%
  \institution{M3S Research Unit, University of Oulu}
  \city{Oulu}
  \country{Finland}
}

\author{Boshuai Ye}
\email{boshuai.ye@oulu.fi}
\authornotemark[1]
\affiliation{%
  \institution{M3S Research Unit, University of Oulu}
  \city{Oulu}
  \country{Finland}
}

\author{Muhammad Azeem Akbar}
\email{azeem.akbar@lut.fi}
\orcid{0000-0002-4906-6495}
\authornotemark[2]
\affiliation{%
  \institution{Software Engineering Department, LUT University}
  \city{53851 Lappeenranta}
  \country{Finland}
}

\author{Javed Ali Khan}
\email{j.a.khan@herts.ac.uk}
\authornotemark[3]
\affiliation{%
  \institution{University of Hertfordshire}
  \city{Hatfield}
  \country{UK}
}
\author{Davoud Mougouei}
\email{davoud.mougouei@deakin.edu.au}
\authornotemark[4]
\affiliation{%
  \institution{Deakin University}
  \city{Geelong}
  \country{Australia}
}

\author{Xinyuan Ma}
\email{xinyuan.ma@student.oulu.fi}
\authornotemark[1]
\affiliation{%
  \institution{M3S Research Unit, University of Oulu}
  \city{Oulu}
  \country{Finland}
}

\renewcommand{\shortauthors}{Khan et al.}

\begin{abstract}

The rise of quantum computing has driven the need for quantum software engineering, yet its programming landscape remains largely unexplored in empirical research. As quantum technologies advance toward industrial adoption, understanding programming aspects is crucial to addressing software development challenges. This study analyzes 6,935 quantum software programming discussion posts from Stack Exchange platforms (Quantum Computing, Stack Overflow, Software Engineering, and Code Review). Using topic modeling and qualitative analysis, we identified key discussion topics, trends (popular and difficult), tools/frameworks, and practitioner challenges. Twenty topics were identified, including popular ones such as \textit{physical theories and mathematical foundations}, as well as \textit{security and encryption algorithms}, while the most difficult were \textit{object-oriented programming} and \textit{parameter control in quantum algorithms}. Additionally, we identified nine frameworks that support quantum programming, with \textit{Qiskit} emerging as the most widely adopted. Our findings also reveal core challenges in quantum software programming, thematically mapped into four areas: \textit{theories and mathematical concepts, algorithms and applications, experimental practices and software development}, and \textit{education and community engagement}. This study provides empirical insights that can inform future research, tool development, and educational efforts, supporting the evolution of the quantum software ecosystem.

\end{abstract}

\ccsdesc[500]{Software and its engineering~Software development techniques}

\keywords{Quantum Computing, Quantum Software Programming, Q\&A Platforms}

\maketitle

\section{Introduction}
Quantum technology is increasingly receiving global attention, bringing various opportunities and advantages across a diverse range of industries, e.g., it extends its transformative impact beyond high-tech domains such as cybersecurity \cite{rawat2023quantum, mosca2018cybersecurity} and aerospace \cite{walthall2022impact} to traditional sectors such as finance and communication \cite{herman2023quantum, yang2023survey}. At the forefront of this revolution is Quantum Computing (QC), a technology driven by the principles of quantum mechanics. QC demonstrates unparalleled capabilities in parallel processing and problem-solving efficiency, making it a key element in this technological shift \cite{steane1998quantum, deutsch1985quantum}. Despite its potential, the pace of quantum software development has been comparatively slow and loaded with challenges \cite{zhao2020quantum}. Overcoming these challenges is critical for QC to fully realize its potential industrial applications \cite{eskandarpour2020quantum, zhao2020quantum, khan2023software}.

The essence of QC lies in its foundation in quantum mechanics, a branch of physics where "quantum" signifies the minimum amount of any physical entity involved in the interaction, such as particles like electrons, neutrons, and photons \cite{messiah2014quantum, steane1998quantum, cross2015quantum, deutsch1985quantum}. Leveraging quantum mechanics characteristics such as superposition, entanglement, and quantum interference, quantum computers outperform classical computers in certain tasks significantly. Quantum computers excel in scenarios requiring high-speed parallel computing, complex algorithmic problem-solving, and advanced computational modeling in fields ranging from astrophysics to climate science \cite{tews2020quantum, ajagekar2022quantum}. They are also adept at executing highly accurate and detailed computations for applications like molecular simulations in drug discovery and optimization of large-scale logical operations \cite{grimsley2019adaptive, mcardle2020quantum,  rebentrost2014quantum}. 

Considering these capabilities, technology giants such as IBM\footnote{\url{https://www.ibm.com/quantum}}, Google\footnote{\url{https://quantumai.google/quantumcomputer}}, and Microsoft\footnote{\url{https://www.microsoft.com/en-us/research/research-area/quantum-computing}} are investing substantially in quantum hardware and software development, aiming to scale QC's applications at industrial level \cite{behera2019designing, courtland2017google}. However, realizing the full potential of quantum computing necessitates the development of effective quantum software solutions, which remains a challenge due to the emerging state of large-scale quantum software development. To address this, the emerging field of Quantum Software Engineering (QSE) seeks to provide developers with the methodologies and tools needed to build quantum applications with the same confidence as their classical counterparts, integrating conventional software engineering practices into the quantum domain \cite{zhao2020quantum}. However, the inherent complexities of quantum mechanics, such as quantum entanglement and superposition, bring significant challenges for developers. These quantum characteristics require not only a new understanding but also novel programming paradigms, which currently lack the robust debugging and error-checking tools that classical programming benefits from \cite{akbar2024quantum, chong2017programming, sofge2008survey}.

Given these challenges, it is essential to understand how developers are currently navigating the complexities of quantum software programming and what difficulties they face in practice. One valuable source of insights into these issues is Q\&A platforms, where developers actively discuss quantum programming concepts, seek solutions to problems, and exchange knowledge. The increasing discussions around quantum programming on Q\&A platforms underscore the growing interest in and challenges of quantum software development. However, the depth and potential insights of practitioners discussions on quantum programming have yet to be fully examined. This study aims to address this gap by investigating the discussions among developers across leading Q\&A platforms, aiming to provide insightful perspectives on quantum software programming that could enrich current and prospective research. To accomplish this and explore these untapped areas, we develop the following research questions (RQs):

\begin{itemize}
    \item \textbf{RQ1:} What are the topics of quantum software programming discussed on Q\&A platforms?
    \begin{itemize}
        \item \textbf{RQ1.1:} What are the most popular and difficult topics related to quantum  programming? 
    \end{itemize}
    \item \textbf{RQ2:} What quantum software programming tools and frameworks have been discussed?
    \item \textbf{RQ3:} What are the challenges of quantum software programming?
\end{itemize}

The aim is to critically evaluate the practitioners discussions, trends, and challenges highlighted by developers on prominent Q\&A platforms, thereby addressing a notable issue in the existing literature. The research findings will highlight the challenges encountered by quantum software developers and contribute a fresh perspective to the quantum software programming dialogue, which is essential for the field development towards real-world application.

The structure of this paper is outlined as follows: Section 2 discusses the research background, Section 3 reviews related work, Section 4 reports the research methodology adopted, Section 5 presents the findings, Section 6 discusses the study implications, Section 7 discusses threats to validity, and Section 8 concludes with future research directions.

\section{Background}
In this section, we present the background concepts related to this study by discussing quantum computing, quantum software engineering, and quantum programming. Understanding these concepts is essential for comprehending the technical environment and the distinct challenges associated with developing quantum technology software. 

\subsection{Quantum Computing}
Quantum computers are based on quantum mechanics concepts to process the data much faster and more efficiently than existing classical computers, though this speedup is limited to specific tasks rather than all computational problems \cite{rieffel2011quantum, deutsch1985quantum, messiah2014quantum, steane1998quantum, cross2015quantum}. Classical computers use bits, which are electrical pulses, to symbolise the binary values of 1 and 0. Quantum computers, in contrast, use qubits, which are subatomic particles like electrons or photons, to indicate the binary values of 1 and 0 \cite{rieffel2011quantum,deutsch1985quantum, messiah2014quantum, steane1998quantum, cross2015quantum}. Qubits, unlike classical bits, can potentially assume the value of 0 or 1 with a certain probability. This characteristic is referred to as the superposition principle \cite{kaye2006introduction}. 

Quantum superposition is a core concept in QC that allows a quantum bit, or qubit, to simultaneously represent multiple states. Specifically, a qubit can be in a state of $|0\rangle$, a state of $|1\rangle$, or any combination of these states \cite{kaye2006introduction}. This ability enables a qubit to exist in a superposition of states, allowing it to effectively explore multiple computational paths simultaneously. This significantly enhances the processing power for certain tasks by leveraging quantum interference to amplify correct outcomes. Mathematically, the state of a qubit can be represented as:

\[
    |\phi\rangle = \alpha |0\rangle + \beta |1\rangle
\]

In this context, $|0\rangle$ and $|1\rangle$ denote the basic states of the qubit, while $\alpha$ and $\beta$ are complex numbers. The absolute values squared of these numbers, $|\alpha|^2$ and $|\beta|^2$, respectively, indicate the probabilities of the qubit collapsing into one of these two states upon measurement, where:

\[
    |\alpha|^2 + |\beta|^2 = 1
\]

Similarly, quantum entanglement refers to a situation in which numerous qubits are strongly connected to the extent that the quantum state of one qubit cannot be described without considering the others \cite{jozsa2003role}. Consequently, when a single qubit is measured, it immediately influences the condition of its entangled counterparts, regardless of the distance in space between them. It was initially introduced by Einstein, Podolsky, and Rosen in 1935 to question the comprehensiveness of quantum mechanics, using what is known as the EPR Paradox \cite{einstein1935can}. However, subsequent research, notably John Bell's formulation of Bell's Theorem in 1964 \cite{bell1964einstein} and Alain Aspect's experimental validations in the early 1980s \cite{aspect1981experimental}, not only validated the presence of quantum entanglement but also exposed it as an inherent and essential feature of quantum mechanics. Both quantum superposition and entanglement are the key principles that allow quantum computers to simultaneously perform many computational routes and facilitate the transfer of complex and correlated information across qubits \cite{kaye2006introduction}.

In QC, the quantum logic gates are the fundamental operational units used to control and manipulate the states of qubits, much like classical logic gates (such as AND, OR, NOT) operate on classical bits \cite{monroe1995demonstration}. Quantum gates apply specific mathematical transformations (typically unitary transformations, which preserve the total probability of quantum states) to alter the superposition states of quantum bits. These transformations can act on a single quantum bit or multiple quantum bits to perform more complex operations, such as the controlled-NOT (CNOT) gate and the Toffoli gate (controlled-controlled-NOT), which enable entanglement and other advanced quantum functionalities among quantum bits \cite{miller2011elementary}. The collection of quantum gates can be assembled into quantum circuits to execute complex quantum algorithms, such as Shor’s algorithm for factoring large integers and Grover's search algorithm \cite{shor1994algorithms, grover1996fast}. Precise design and implementation of quantum gates are among the key challenges in QC technology, as they must be meticulously controlled to avoid quantum decoherence and other quantum noise effects.

QC offers substantial advantages over classical computing regarding processing speed and efficiency. For instance, Shor's algorithm can factorize large integers much faster than any known classical algorithm \cite{bhatia2020efficient}. QC can efficiently and accurately simulate intricate systems, such as molecular structures or optimization and simulation of industrial processes \cite{childs2018toward, mcardle2020quantum, rebentrost2014quantum}. Another promising application lies in machine learning \cite{biamonte2017quantum}, where quantum algorithms potentially provide accelerated data processing capabilities. As the field matures, future directions encompass the expansion of quantum systems-enhancement of qubit coherence and the establishment of resilient quantum error correction techniques \cite{jayashankar2023quantum, knill1998resilient}. This technological progress will enhance the accessibility and feasibility of QC, enabling it to address a wider range of intricate computational issues.

\subsection{Quantum Software Engineering}
Quantum software engineering (QSE) is an emerging research domain that addresses the challenges of developing and maintaining quantum software applications. Zhao \cite{zhao2020quantum} presented the concept of quantum software engineering and defined the entire QSE life cycle, including requirements analysis, design, implementation, testing, and maintenance. Zhao conducted a detailed literature study of existing QSE practices and concluded that classical software engineering practices must be adopted to develop quantum software. He further highlighted that QSE could only be actualize based on the concepts of hybrid computing by integrating the classical and quantum computing practices \cite{zhao2020quantum}. Similarly, this is highlighted in the national agenda for software engineering research and development
presented by SEI-Carnegie Mellon University \cite{carleton2022architecting}: \textit{"Advances in software engineering for quantum are as important as the hardware advances. The goals of this research area are to first enable current quantum computers to be programmed more easily and reliably, and then enable increasing abstraction as larger, fully fault-tolerant quantum computing systems become available. A key challenge is to, eventually, fully integrate these types of systems into a unified classical and quantum software development lifecycle."}

QSE faces significant challenges primarily due to the lack of tools, processes and solutions available for quantum software development activities. In contrast to well-established tools and frameworks in classical software development, quantum software tools are still being developed \cite{junior2021systematic}. This lack of mature tools complicates ensuring the reliability and correctness of quantum programs, which is crucial given the probabilistic nature of quantum outcomes. Errors, even minor, can lead to significant deviations in the results. Another challenge is scalability. The theoretical exponential speed-ups promised by QC will only materialize if quantum software can effectively scale with increasing qubits \cite{murali2019noise}. Today’s quantum software practices must evolve to handle scalability and integration issues, particularly how quantum and classical systems work together in practical applications \cite{oskin2002practical}. 

The QC community is pursuing several research avenues in response to these challenges. Enhancing and creating new quantum simulators is one approach aiming to provide more accurate and efficient simulations of quantum algorithms on classical systems \cite{aspuru2012photonic}. These tools are indispensable for developing quantum software without extensive quantum hardware. Another area of focus is developing a lifecycle management framework for quantum software, adapting the best practices from classical software engineering to suit the quantum context \cite{weder2020quantum}. This framework would cover all aspects, from design and development to testing and maintenance. There is also a growing interest in developing programming models that integrate quantum and classical computing processes seamlessly \cite{aithal2023advances}. This hybrid approach is crucial for building practical applications that harness the advantages of QC while still relying on classical computations for certain tasks. As QSE continues to evolve, its role in transforming the theoretical benefits of QC into practical, reliable, and scalable software applications becomes increasingly significant. This progress will expand quantum computer capabilities and broaden their applicability across various sectors, leading to innovative solutions for complex challenges.

\subsection{Quantum Programming}
Quantum programming is used to build and design quantum software solutions and algorithms \cite{selinger2004towards}. It involves utilizing specific programming languages and tools to manage and manipulate qubits on quantum computers. The key differences between quantum programming and classical programming are the fundamental concepts of quantum mechanics, such as quantum superposition and entanglement. Quantum programming is essential for realizing the real-world implications of QC by developing software applications that perform a particular task that is classically hard to achieve \cite{johnston2019programming}. 

Various quantum programming tools and frameworks have been developed and are used to create quantum applications. For instance, Qiskit is an open-source QC framework developed by IBM \cite{ibm2021qiskit}. It is used for the development, simulation, and execution of quantum applications; Cirq, developed by Google is a software development framework that specifically emphasizes defining quantum circuits at low level, specifying individual gates and operations \cite{google2022cirq}. QuTiP is a powerful and widely used open-source Python library for simulating quantum systems, mostly used for conducting numerical simulations in the field of quantum information processing \cite{nation2011qutip}. Similarly, Microsoft Q\# \cite{microsoftQ} is a specialized high-level programming language designed for quantum computing, integrated within the Microsoft Quantum Development Kit. It focuses on the modularity and testability of quantum programs, allowing developers to write and debug quantum algorithms efficiently. Since quantum computers have not yet become widely available, most quantum programs currently necessitate simulation and testing on classical computers. This simulation validates the accuracy and effectiveness of quantum algorithms and forecasts the behaviour of quantum programs. Quantum programming is a rapidly developing domain. The demand for quantum programmers increases as technology progresses and more quantum gear becomes accessible. Acquiring proficiency in quantum programming necessitates comprehension of fundamental concepts of quantum mechanics and expertise in specialized programming skills and tools to effectively create and execute quantum algorithms \cite{de2022software}.

\section{Related Work}
To the best of our knowledge, this is the first comprehensive empirical study (mining multiple repositories) that focused on quantum software programming, consisting of programming topics, popular and difficult topics, programming tools and frameworks, and challenges of quantum software programming. This section explores related work addressing various studies focusing on quantum software engineering and quantum software programming \cite{li2021understanding, de2022software, aktar2025architecture, piattini2020talavera, zhao2023empirical, ali2022software, zhao2020quantum, heim2020quantum}.

Li et al. \cite{li2021understanding} conducted an empirical study on Q\&A platforms, which has investigated the dynamics of community interactions and the nature of the challenges discussed within the QSE domain. This study provides an analytical overview of the questions and issues raised by developers engaged in QSE, offering insights into the most frequent stumbling blocks encountered in the field. This work focuses on identifying common challenges faced by quantum software developers and highlights unique challenges of QSE, such as interpreting quantum program outputs, understanding quantum algorithm complexity, and bridging the knowledge gap between quantum and classical computing. The study also identified critical areas requiring future attention, including dependency management, environment setup, and learning resources for quantum programming.

De Stefano et al. \cite{de2022software} conducted an exploratory study on quantum programming by analyzing GitHub repositories and surveying developers. The study aims to identify the purpose of quantum programming frameworks and the challenges of quantum software programming. The study highlights the need for comprehensive tools and methodologies to aid quantum developers. A taxonomy of challenges was provided, covering software infrastructure, hardware, and program comprehension. This work offers valuable insights into the challenges faced by quantum developers. It underscores the importance of advancing tools and resources for QSE.

Aktar et al. \cite{aktar2025architecture} conducted an empirical study to investigate architectural decisions in quantum software systems, using data from GitHub and Stack Exchange. The study identified six linguistic patterns used in architecture decision-making, with "Solution Proposal" and "Information Giving" being the most frequent. Key decision categories include implementation and technology decisions, while maintainability and performance were highlighted as critical quality attributes. The authors also identified design issues and high error rates as significant challenges faced by practitioners. This work provides valuable insights into the complexities of architectural decision-making in quantum software systems, emphasizing the need for tailored methodologies and tools to address these challenges.

Zhao et al. \cite{zhao2023empirical} examined 391 software bugs obtained from 22 open-source quantum frameworks related to quantum machine learning (QML). The study findings indicate that 28\% of the bugs are associated with QML, mainly related to qubit manipulation, kernel, drawer, template, interface, algorithm, transform, and operation. Additionally, the authors provides an overview of the primary challenges encountered by QML framework developers. 

The "Talavera Manifesto" \cite{piattini2020talavera} presented at the first International Workshop on Quantum Software Engineering in 2020 outlines a set of fundamental principles for the emerging discipline of QSE. This manifesto consists of a set of fundamental principles for the growing QSE field, developed collaboratively by researchers and practitioners. It outlines essential principles for QSE, emphasizing technological neutrality to enhance field adaptability and the integration of classical and quantum computing. It highlights crucial elements such as project management, zero faults for reliability, quality assurance, and software reuse. Additionally, it stresses the importance of rigorous security, privacy measures, and robust governance to guide the development and implementation of quantum technologies, aiming to discipline and align QSE with industry standards and technical progress. 

Ali et al. \cite{ali2022software} emphasize the critical role of QSE in enabling the development of cost-effective and scalable quantum applications. Their work outlines the need for a robust quantum software stack, ranging from operating systems to programming languages, to address the unique challenges posed by quantum computing, such as superposition and entanglement. They highlight the lack of well-defined Software Development Life Cycles (SDLCs) for quantum software and underscore the importance of novel methodologies for requirements engineering, modeling, testing, and debugging. The study also identifies significant opportunities for advancing quantum software tools and frameworks, which are essential for realizing the full potential of quantum technologies in domains such as healthcare, finance, and physics.

Furthermore, Zhao \cite{zhao2020quantum} conducted an extensive literature review of quantum software engineering, accurately examining each phase of the quantum software development lifecycle, including requirements analysis, software design,  implementation, and testing. This comprehensive review provided valuable insights into the systematic processes involved in developing quantum software, which remains an emerging field with complex technological nuances. In addition to describing these phases, Zhao highlighted the theoretical aspects supporting quantum computing technologies, such as quantum state management, entanglement, and superposition, which directly influence software design and functionality. The review also focused on the emerging trends and potential challenges that quantum software developers might face, such as scalability issues, quantum hardware limitations, and the need for new debugging techniques specific to quantum software programming. Furthermore, Zhao explored the future prospects of quantum software development, suggesting areas where quantum computing could offer transformative capabilities, including cryptography, complex system simulation, and optimization problems.

Similarly, Heim et al. \cite{heim2020quantum} provides an overview of quantum software programming languages. The authors review several state-of-the-art quantum programming languages, highlighting their design priorities, unique features, and applications across various user groups, from beginners to seasoned practitioners. Their work underscores the evolving ecosystem of quantum programming languages and its impact on advancing both theoretical and practical quantum computing.
\subsection{Comparative Analysis}
We now compared our work with existing related studies based on the above discussions (See Table \ref{tab:comparison}). In reviewing the current literature, several theoretical investigations and initial empirical studies have been conducted in the field of quantum software programming. However, this research study is significantly distant from the existing related work studies. We noticed that there are still significant gaps in understanding the real-time, practical challenges that developers encounter in this emerging area. For instance, the study conducted by Li et al. \cite{li2021understanding} explored Q\&A repositories to examine the challenges and opportunities in quantum software engineering. This was one of the earliest studies to mine repositories in this domain. However, its focus was primarily on the overall QSE lifecycle as perceived by practitioners. While the study briefly touched on quantum software programming topics, it lacked in-depth insights into the quantum software programming domain. 

The study conducted by De Stefano et al. \cite{de2022software} is closely related to our work. The authors primarily explored GitHub repositories to understand the current state of quantum programming and the challenges faced by developers, particularly in relation to using quantum programming frameworks. However, their study did not provide a comprehensive overview of quantum software programming, including diverse aspects such as quantum programming topics, popular and difficult topics, and broader overview of programming tools, frameworks as addressed in this study. 

On the other hand, Aktar et al. \cite{aktar2025architecture} mined StackExchange and GitHub repositories to investigate the limitations and challenges associated with architectural decisions in quantum software development. However, the study did not specifically address aspects related to quantum software programming. Similarly, Zhao et al. \cite{zhao2023empirical} focus narrowly on bugs within quantum machine learning platforms and categorize them based on symptoms and root causes. 

In studies such as Ali et al. \cite{ali2022software} and Piattini et al. \cite{piattini2020talavera}, foundational insights into QSE were provided by defining principles, commitments, and approaches for hybrid (quantum and classical) software engineering, as well as addressing the challenges of hybridization. However, neither of these studies conducted an empirical investigation into quantum software programming. Similarly, the review studies by Heim et al. \cite{heim2020quantum} and Zhao et al. \cite{zhao2020quantum} briefly focus on quantum programming languages and the broader landscape of Quantum Software Engineering. While these studies provide valuable theoretical and conceptual insights, they do not include empirical investigations or address the practical challenges and real-world experiences of quantum programming.

\begin{table}[ht]
\centering
\caption{Comparison Summary. Note: (\checkmark: included, \xmark: not included, *: abstract overview)}
\resizebox{\textwidth}{!}{%
\begin{tabular}{|p{3.5cm}|c|c|c|c|c|}
\hline
\textbf{Studies} & \textbf{Quantum Software Programming} & \textbf{Repositories Mining} & \textbf{Topics Popularity and Difficulty} & \textbf{Tools and Frameworks} & \textbf{Challenges} \\ \hline
Li et al. \cite{li2021understanding} & \cmark (*) & \cmark & \cmark & \cmark (*) & \cmark \\ \hline
De Stefano et al. \cite{de2022software} & \cmark & \cmark & \xmark & \cmark (*) & \cmark \\ \hline
Aktar et al. \cite{aktar2025architecture} & \xmark & \cmark & \xmark & \xmark & \xmark \\ \hline
Zhao et al. \cite{zhao2023empirical} & \cmark (*) & \cmark & \xmark & \xmark & \cmark \\ \hline
Ali et al. \cite{ali2022software} & \cmark (*) & \xmark & \xmark & \cmark (*) & \cmark (*) \\ \hline
Piattini et al. \cite{piattini2020talavera} & \cmark (*) & \xmark & \xmark & \xmark & \xmark \\ \hline
Heim et al. \cite{heim2020quantum} & \cmark & \xmark & \xmark & \cmark & \cmark (*) \\ \hline
Zhao et al. \cite{zhao2020quantum} & \cmark (*) & \xmark & \xmark & \cmark (*) & \cmark (*) \\ \hline
Proposed Study & \cmark & \cmark & \cmark & \cmark & \cmark \\ \hline
\end{tabular}%
}

\label{tab:comparison}
\end{table}

\section{Research Methodology}

\subsection{Research Questions} \label{Sec:Research Questions}
This research study aims to investigate the discussion regarding quantum software programming by analysing the data collected from several Q\&A platforms. The research questions (RQs) are particularly designed to reveal crucial insights into the common topics, tools, frameworks, and challenges in the quantum software programming domain. To answer these RQs, 6935 posts relating to quantum software programming are retrieved from the selected Q\&A repositories (See Section \ref{Data Collection}) based on the methodology outlined as follows:

\textit{\textbf{RQ1}: What are the topics of quantum software programming discussed on Q\&A platforms?}

\textbf{Rationale}: The first research question (RQ1) aims to identify and categorise discussion topics on quantum software programming in various Q\&A platforms. It will provide insights into the primary areas of interest within the field and help the software engineering community to better understand the thoughts and needs of QC developers. 

\textbf{Method}: Following the research conducted in \cite{zahedi2020mining, barua2014developers, haque2020challenges}, we used the Latent Dirichlet Allocation (LDA) \cite{blei2003latent} technique, a well-known unsupervised learning NLP approach, for conducting topic modeling on the data obtained from the selected Q\&A platforms  \cite{chen2016survey}. 

The LDA approach identifies topics in a collection of documents by grouping similar words together into topics based on their distribution across the documents. For instance, Zahedi et al. \cite{zahedi2020mining} applied the LDA approach to analyze discussion topics on Stack Overflow (SO) and identify popular topics in the continuous software engineering domain. Similarly, Barua et al. \cite{barua2014developers} examined SO and adopted the LDA approach to uncover trends, topic relationships, and evolving developer interests over time. In another study, Haque et al. \cite{haque2020challenges} analyzed the SO repository using the LDA technique to classify popular topics in Docker-related Q\&A discussions and identify key challenges faced by developers.

Similarly, in this study, each post—including the title, body, and answers—is considered a unified document, and the LDA approach is used to examine word frequency to determine the primary topics discussed in quantum software programming. The LDA model will be configured to analyze and categorize words into topics by examining their co-occurrence patterns across the documents. It works on the assumption that each document is a mixture of topics and that each topic is a distribution of words. The key parameters, including the number of topics and the distribution of words within each topic, will be optimized iteratively to enhance model performance and coherence scores. Through statistical analysis of word distributions, the LDA model identifies groups of words that frequently appear together, uncovering latent topics underlying the text corpus. This approach not only uncovers the primary topics, but also highlights the distinctiveness of the discussions around these topics within the community. The core subject represented by each identified topic will be analyzed, and the most representative terms within each topic cluster will be used to assign accurate and meaningful labels to the topics. This labeling process aids in summarizing and effectively conveying the essential elements of the discussions in a clear and concise manner. The results of the LDA analysis will offer a structured and comprehensive overview of the discussions within the quantum software programming domain. This synthesis will highlight and categorize the most prevalent topics, providing valuable insights into the community primary focus areas, and addressing RQ1 about the primary topics discussed on Q\&A platforms in the quantum software programming domain.

\textit{\textbf{RQ1.1}: What are the most popular and difficult topics?}

\textbf{Rationale}: Identifying popular and difficult topics enables to evaluate community interest and the complexity of quantum software programming. By analysing the popularity and difficulty of the topics discussed, the study can highlight where the community is most engaged and identify areas that may require additional research, better resources, or enhanced instructional materials to solve current challenges. 

We adopted the concepts of topic popularity and difficulty from previous repostries mining studies in software engineering domain \cite{zahedi2020mining, Peruma_2021, bagherzadeh2019going, abdellatif2020challenges}. The popularity reflects the discussion topics that are widely adopted or frequently discussed by developers. Understanding these topics/trends helps prioritize resources, such as targeted training programs or tool development, to address the significance of these topics. Additionally, it provides valuable insights for future research and development, ensuring that efforts focus on improving and advancing the most relevant areas.
Examining topic difficulty helps evaluate the challenges associated with responding to questions related to different topics. Identifying topics with fewer responses highlights areas where greater community engagement is needed. Furthermore, this analysis emphasizes the need for better frameworks or tools to assist developers in effectively addressing these challenging areas.

It is important to distinguish between topic difficulty and challenges within a topic. Challenges refer to the complexities faced in a particular area, topic difficulty specifically refers to how hard it is for developers to provide answers to related questions.

\textbf{Method}: To assess the popularity and difficulty of topics in the quantum software programming domain, a dual metric approach referenced in studies \cite{ahmed2018concurrency, le2021large, rosen2016mobile} has been used. It includes a combination of interaction metrics for popularity—views, score, favorite count, and comments—and response metrics for difficulty—the percentage of accepted answers, median duration to an accepted answer, and the ratio of answers to views. Popularity for each topic is calculated using the geometric mean of the respective metrics, including P1 (Views), indicating the level of interest a topic garners; P2 (Score), reflecting community approval through upvotes minus downvotes; P3 (Favourite Count), showing how many times users have bookmarked a topic as valuable; and P4 (Comments), indicating the level of engagement and discussion depth a topic generates. While difficulty is determined by D1 (Percentage of Accepted Answers), which inversely indicates difficulty by showing the proportion of questions that receive satisfactory answers; D2 (Median Duration to Accepted Answer), which directly correlates with difficulty as it measures how long it takes for questions to receive an accepted answer; and D3 (Answer to Views Ratio), where lower ratios suggesting higher difficulty due to fewer responses per view, implying the topic might be more complex or less understood. The reciprocals of D1 and D3 are used to ensure the scale aligns correctly, where higher values signify greater difficulty. Next, the geometric mean (Equation 3) for both Popularity (Equation 1) and difficulty (Equation 2) have been calculated. To normalize these metrics across topics, ensuring comparability, we applied min-max normalization. It adjusts each topic's popularity and difficulty index to a scale from 0 to 1, calculated as: (Equation 4) and (Equation 5). This normalization helps in reducing the bias of outlier values and balances the scale of measurements. Finally, the normalized scores are visualized through bar graphs, providing a clear, comparative view of which topics are most \textit{popular} and which are perceived as difficult. This approach not only highlights the areas with high \textit{popularity} and difficulty but also identifies potential gaps in knowledge or resources that can benefit from additional attention.

\begin{align}
    \text{Popularity}_i &= G(P1_i, P2_i, P3_i, P4_i) \tag{1} \\
    \text{Difficulty}_i &= G\left(\frac{1}{D1_i}, D2_i, \frac{1}{D3_i}\right) \tag{2} \\
    G(x_1, x_2, \ldots, x_n) &= \left( \prod_{i=1}^{n} x_i \right)^{\frac{1}{n}} \tag{3} \\
    \text{Normalized}_{\text{Popularity}} &= \frac{\text{Popularity}_i - \min(\text{Popularity})}{\max(\text{Popularity}) - \min(\text{Popularity})} \tag{4} \\
    \text{Normalized}_{\text{Difficulty}} &= \frac{\text{Difficulty}_i - \min(\text{Difficulty})}{\max(\text{Difficulty}) - \min(\text{Difficulty})} \tag{5}
\end{align}

\textit{\textbf{RQ2}: What quantum software programming tools and frameworks have been discussed?}

\textbf{Rationale}: The discussion around tools and frameworks in quantum software programming is important as they serve as the practical instruments by which concepts and theories are implemented in real-world situations. This question aims to identify and describe the tools and frameworks currently used in the field, providing insight into the technological environment of quantum software programming. To enhance their ability to navigate the ecosystem, academics and practitioners can gain valuable insights by familiarising themselves with the existing and widely adopted tools. This knowledge can help them identify areas that require further research or innovation in the domain of tools and frameworks.

\textbf{Method}:   
To address RQ2, the data extracted from the selected Q\&A repositories was systematically analyzed to identify quantum software programming tools and frameworks. A comprehensive set of keywords and tags was developed to guide this analysis. Specific tool names such as "Qiskit," "Cirq," "Braket," and "Strawberry Fields" were included alongside broader terms including "quantum SDK," "quantum simulator," "quantum programming library and "quantum programming framework". These keywords were carefully selected based on an background related studies \cite{zhao2020quantum, heim2020quantum, de2022software, sofge2008survey} and common terminology used in the quantum software programming community.

The search results undergo manual filtering to ensure their relevance by excluding posts that do not substantially address the application of the tools in quantum programming. The analysis concentrate on succinctly summarising the functionalities and identifying the limitations of each tool based on community comments. Understanding the practical usability of these tools and frameworks is essential for evaluating their performance and how well they meet the needs of developers. 

\textit{\textbf{RQ3}: What are the challenges of quantum software programming?}

\textbf{Rationale}: Quantum software programming is a rapidly evolving field of research that presents unique challenges arising because of the novice field of quantum software engineering, the complexity of quantum mechanics, and the existing limitations in quantum hardware. It is essential to identify these challenges as they bring attention to the actual difficulties that practitioners face in real-world scenarios, such as problems related to the development and scalability of quantum software, comprehending quantum mechanics, or the absence of robust programming frameworks and tools. By recognising these challenges, the research can make valuable contributions to focused advancements in QC education, tool development, and practical application frameworks.

\textbf{Method}: For RQ 3, which aims to uncover the critical challenges in quantum software programming, a mixed-methods approach has been used, combining both \textit{quantitative} and \textit{qualitative} analyses to provide a comprehensive understanding.

\begin{itemize}
    \item \textbf{Quantitative Analysis} - Initially used the Accumulated Post Score (AMS) formula (\textit{see Equation 6}) proposed by Bajaj et al. \cite{bajaj2014mining}. It is a metric used to measure the overall engagement and relevance of posts on Q\&A platforms.

    \[
    AMS_i = 3S_i + 10C_i + A_i + F_i \tag{6}
    \]

    In Equation 6, $S_i$, $C_i$, $A_i$, and $F_i$ denote the Score, Comment count, Answer count, and Favourite count of the $i$-th post respectively. This formula helps quantify the \textit{practitioners' engagement} and the relevance of discussions on Q\&A platforms. Through the calculation of the AMS metric proposed by Bajaj et al. \cite{bajaj2014mining} for all relevant posts, we identified and selected the top 200 posts based on their accumulated post scores for a comprehensive qualitative examination. 

    This selection criterion was chosen because posts with higher scores are likely to represent the most impactful and resonant discussions. Upon reviewing these 200 posts, we observed recurring insights, confirming that this subset was rich in information and representative of the overall dataset. Therefore, the selected posts offer a comprehensive and representative sample for examining the key challenges practitioners face in quantum software programming.

    \item \textbf{Qualitative Analysis} - For the qualitative data analysis, the thematic analysis approach defined by Braun and Clarke \cite{braun2006using} has been adopted, using NVivo\footnote{\url{https://lumivero.com/products/nvivo/}}. The initial stage involved a methodical categorization of the selected 200 posts to extract initial \textit{characteristics} that represent the challenges being discussed. The coding process required the manual and independent identification and categorization of common concepts and concerns into initial codes. These codes were subsequently refined and grouped into broader thematic categories. The emergent themes were further structured into sub-themes to elaborate more detail on specific challenges in quantum programming.

 The first, second, and last authors jointly participated in data labeling, coding, and mapping processes to address data differences. In cases where the lead authors did not agree, the final conclusions were made based on the consensus reached by all authors to improve the reliability of the results. This streamlined approach not only highlights the common challenges but also offers a framework for investigating possible solutions in the domain of quantum software programming.
\end{itemize}

\begin{figure}
    \centering
    \includegraphics[width=1.0\linewidth]{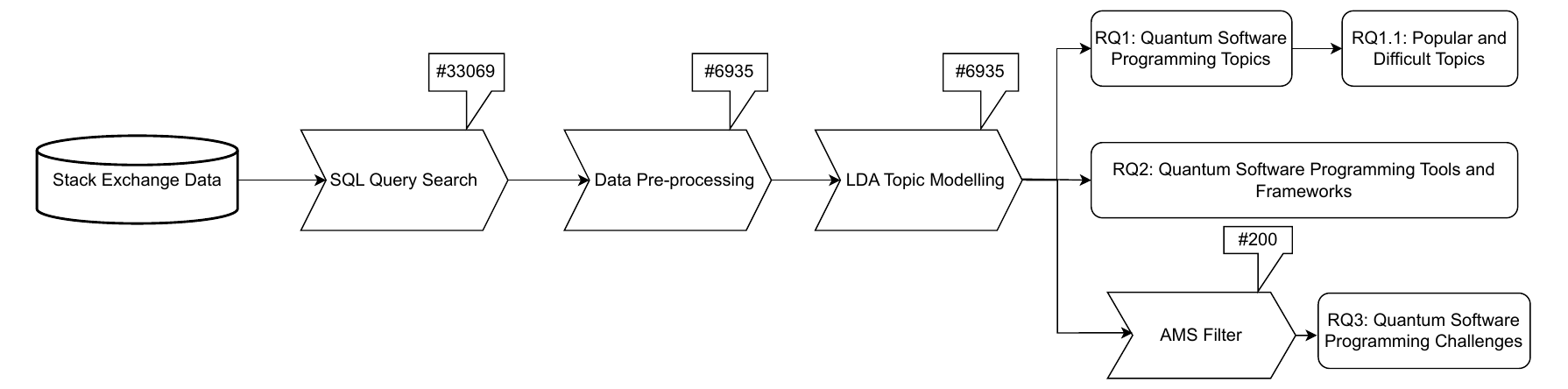}
    \caption{The search, data extraction and filtration process (with the number of posts returned after the step}
    \label{fig:procudure}
\end{figure}

\subsection{Data Collection} \label{Data Collection}

We collected 6935 developers' discussion posts from various Stake Exchange platforms using the Stack Exchange Data Explorer\footnote{\url{https://data.stackexchange.com/}} (SEDE), a robust tool that provides access to the latest "data dumps" of publicly available content. Through the execution of SQL queries, we methodically extracted important data in two primary phases. Initially, data collection focused on Stack Overflow\footnote{\url{https://stackoverflow.com/}} (SO), the most extensive site within the Stack Exchange. Since its inception, SO has transformed from a simple question-and-answer platform into a vast repository of developer knowledge, covering a broad spectrum of topics, from general discussions to technical issues, including those on quantum software programming. Further data was collected from other specialized Stack Exchange sites, including the Software Engineering\footnote{\url{https://softwareengineering.stackexchange.com/}} (SE) site,  addresses broader queries related to the software development lifecycle and is geared towards developers seeking generalized guidance. Additionally, we extracted data from the Code Review\footnote{\url{https://codereview.stackexchange.com/}} (CR) site, aimed at peer-review enhancements of existing code, and the Quantum Computing\footnote{\url{https://quantumcomputing.stackexchange.com/}} (QC) site, dedicated to the specialized field of quantum computing.

This diversity in focus and audience across the sites provides a rich dataset for analysis. Overall SO, being the most prominent, attracts a vast number of users, significantly more than the other sites. Basic statistics from all four sites are presented in Table \ref{tab:stats} to provide insights into the varying dynamics and user engagement across these platforms. We generally categorized the number of questions asked across the selected Q\&A repositories, along with the corresponding answers (count and percentage) and the users involved. It provides insights into question frequency, answer rates, and user engagement across the selected Q\&A repositories. It is identified that SO is by far the largest, with millions of questions and answers, reflecting its wide user base and extensive topic coverage including software engineering practices and methodologies \cite{khan2024insights}.  The SE site has a high percentage of answered questions, indicating active participation and a high response rate. CR also shows significant engagement, particularly in peer-reviewed contexts. The QC site, though smaller, focuses on a specialized field, with a considerable amount of user interaction relative to its size.

Overall, out of 6935 posts,  44 posts are collected from Stack Overflow (SO), 25 from Software Engineering (SE), 6841 from Quantum Computing (QC), and 25 from Code Review (CR). The significantly higher number of posts in the QC repository highlights its focused nature and the community's active interest in quantum-related topics. In contrast, repositories like SO, SE, and CR, which cater to broader and more general technical audiences, have fewer posts specifically relevant to quantum computing, as their scope includes a wide variety of programming and engineering topics rather than being exclusively dedicated to quantum technologies.
The complete data replication package is provided in Khan et al. \cite{khan_2024_14312165}, and the data were retrieved on 23 April 2024.

\begin{table}[h!]
\centering
\caption{Sites basic statistics as of May of 2024}
\label{tab:stats}
\begin{tabular}{|l|l|l|l|l|}
\hline
\textbf{Repositories} & \textbf{Questions} & \textbf{Answers} & \textbf{Answered} & \textbf{Users} \\ \hline
SO                   & 24m (million)      & 36m              & 70\%              & 23m            \\ \hline
SE                   & 64k (thousand)     & 179k             & 95\%              & 376k           \\ \hline
CR                   & 78k                & 122k             & 92\%              & 242k           \\ \hline
QC                   & 11k                & 13k              & 80\%              & 23k            \\ \hline
\end{tabular}
\end{table}

In the Stack Exchange Data Explorer (SEDE), questions and responses are both categorised as posts and stored in the Posts table. SQL query was designed to retrieve only questions, which are identified by a \texttt{PostTypeId} value of 1, while replies are assigned a value of 2. This query was specifically designed to deal with the questions posed by participants in Q\&A platforms \cite{yang2016security}. At first, we used tags to sort the posts. These tags are intended to help users and researchers easily identify and locate questions related to specific topics.

Nevertheless, this strategy was found to be less than perfect for many reasons. Users frequently fail to correctly classify their inquiries using appropriate tags; some include several tags to enhance the understanding and increase the probability of getting responses, even if such categories do not necessarily correspond to the topics of their posts \cite{bajaj2014mining, barua2014developers, dietrich2019man}. Although tags can assist in identifying quantum software programming questions, focusing exclusively on them may result in overlooking essential discussions relevant to the topic in question content. For this purpose, the developer posts selection criteria are broadened by utilising a collection of keywords discovered from previous studies, including (quantum-computing, quantum-programming, quantum-algorithms, quantum-machine-learning, quantum-error-correctional, quantum-software, quantum-code, and quantum-hardware) \cite{altenkirch2005functional, heim2020quantum, selinger2004towards, omer2005classical}. These keywords facilitated the efficient collection of postings from selected Stack Exchange sites. The details of the SQL Query in this research are given in Table \ref{tab:sql_query}, which provides a list of fields and their descriptions for an SQL query. It includes identifiers, question metadata (such as title, body, view count, and score), engagement metrics (upvotes, downvotes, comments, answers), and details about the accepted answer, including its ID and acceptance date. In addition, a duplicate check was incorporated to remove posts that were repeated on multiple platforms. Overall, this methodology resulted in a total of 6935 posts relevant to quantum software programming, along with the corresponding information for these posts.

\begin{table}[h!]
\centering
\caption{Fields and Descriptions for SQL Query}
\label{tab:sql_query}
\begin{tabular}{|l|p{10cm}|}
\hline
\textbf{Field Name}        & \textbf{Description} \\ \hline
\textbf{Id}                & Unique identifier of the question. \\ \hline
\textbf{QuestionURL}       & URL of the question, constructed by appending the Id to the base URL \texttt{https://stackoverflow.com/q/}. \\ \hline
\textbf{Title}             & Title of the question. \\ \hline
\textbf{Body}              & Body text of the question. \\ \hline
\textbf{ViewCount}         & Number of times the question has been viewed. \\ \hline
\textbf{Score}             & Score of the question, typically the difference between upvotes and downvotes. \\ \hline
\textbf{UpVotes}           & Number of upvotes the question has received. This is determined by a subquery counting votes of type 2 associated with the question. \\ \hline
\textbf{DownVotes}         & Number of downvotes the question has received. This is determined by a subquery counting votes of type 3 associated with the question. \\ \hline
\textbf{CommentCount}      & Number of comments on the question. \\ \hline
\textbf{AnswerCount}       & Number of answers to the question. \\ \hline
\textbf{FavoriteCount}     & Number of times the question has been marked as favorite. \\ \hline
\textbf{AcceptedAnswerId}  & Identifier of the accepted answer to the question (if any). \\ \hline
\textbf{AcceptedAnswerDate}& Creation date of the accepted answer. \\ \hline
\end{tabular}
\end{table}

\subsection{Topic Modelling} \label{Topic Modelling}
Topic Modelling is a well-known Natural Language Processing (NLP) technique that structures information, such as topics, from a set of documents. Researchers can use it to identify the core theme structures in an extensive array of texts, uncovering patterns of word clusters and topic distributions without the need to manually analyse each document \cite{chen2016survey}. In topic modeling, Latent Dirichlet Allocation (LDA) \cite{blei2003latent} is the most common and effective unsupervised algorithm, mainly used to extract topics from a textual corpus. It has been successfully used in many different domains, such as continuous software engineering  \cite{zahedi2020mining}, blockchain platforms \cite{wan2019programmers}, software methodologies and practices identification \cite{khan2024insights},  challenges in docker development \cite{haque2020challenges} financial market analysis \cite{yono2020construction}, social media monitoring \cite{wang2012tm}, and the analysis of customer Feedback \cite{wang2018topic}. The LDA algorithm is based on the idea that documents are composed of a mixture of topics, where each topic is characterized by a distribution of words. Thus, words or phrases that appear frequently in the text data can indicate a particular topic \cite{chen2016survey}. This method not only facilitates the process of associating text with important topics but also functions as a powerful tool to conduct exploratory data analysis, thereby enhancing the comprehensiveness and scope of informational insights obtained from extensive datasets. Following the prior research conducted by Mansooreh et al. \cite{zahedi2020mining}, we define the input document as consisting of a question’s title, body, and related responses. This document is then used to provide input to the LDA model, which identifies and outputs the most frequently recurring topics within the text corpus. 

\subsubsection{Pre-processing of collected posts}
Before applying the LDA approach, it is important to pre-process the data to enhance the quality of the results. This pre-processing involved several steps:

\begin{itemize}
    \item \textbf{Removal of Noise:} Stack Exchange uses HTML to format their content, raw text data may contain HTML tags. Removing these tags is essential as they do not contribute to understanding the text content \cite{khan2024insights}. Additionally, punctuation and code snippets that do not enhance the quality of the written content are also excluded. Code snippets, although commonly used in technical discussions, do not contribute to topic modeling since they do not reflect any themes in natural language. For this purpose, the NLTK\footnote{\url{  https://github.com/nltk/nltk}} stop-words corpus is used to remove common terms like ``and'' and ``the,'' which are known as stop-words. These words are often present in all texts and do not contribute to the themes being modeled, therefore they are deleted to clean the textual data.
    
    \item \textbf{Normalization:} The whole content was transformed to lowercase to maintain uniformity throughout the dataset. It ensures that words with different cases are not regarded as separate tokens, hence it improves homogeneity. (e.g., ``Apple'' vs. ``apple'').
    
    \item \textbf{Text Simplification:} Next, both lemmatization and stemming methodologies are used to streamline words to their essential or core forms, thereby decreasing grammatical variability within the text. Lemmatization is a linguistic process that converts words to their lemma form, taking into consideration the context and part of a sentence. For example, it transforms the word ``running'' into ``run.'' Stemming is a procedure that reduces words by removing the endings, resulting in words like ``connection,'' ``connected,'' and ``connecting'' being reduced to ``connect.''
    
    \item \textbf{Word Frequency Filtering:} According to the approach outlined by Yang et al. \cite{yang2016security}, the words that have a frequency of less than 10 times in the dataset are removed. By implementing this filtering approach, the model focuses on terms that have more impact, resulting in a smaller vocabulary size. This approach can lead to faster processing speed and improved model quality.
\end{itemize}

\subsubsection{Topic modelling with LDA}

In this study, after pre-processing the dataset, the LDA algorithm has been applied to analyse the collected textual data and unveil the thematic structures based on the discussions of quantum software programming across the selected repositories. It began by determining the optimal number of topics $K$, aiming to strike a balance between overly broad and excessively detailed topics. Setting $K$ too low risks overlapping topics that are hard to generalize, while too high a $K$ might fragment the topics excessively, diluting their interpretability. In this study, the $K$ values range from 1 to 50 are calculated. Alongside $K$, the Dirichlet hyperparameters, alpha ($\alpha$) and beta ($\beta$) values are also calculated, where $\alpha$ influences the number of topics per document, and $\beta$ affects the distribution of words within the topics. We tested values from 0.01 to 1, adjusting in increments of 0.2 to explore both symmetric and asymmetric distributions across the documents, as suggested in previous research \cite{ahmed2018concurrency, barua2014developers, le2021large, rosen2016mobile}.

For constructing the LDA model, the Gensim\footnote{\url{  https://github.com/piskvorky/gensim}} library is used, formatting the pre-processed texts as bags of words. This format is essential for the LDA algorithm to assign probabilistic topic distributions effectively. Batch learning is selected to ensure consistent results across different model runs, which is crucial for accurate model comparisons. The effectiveness of each model setup was evaluated based on the Coherence Score, which measures the semantic similarity of the high probability words within each topic, thus ensuring their interpretability \cite{abdellatif2020challenges, rosen2016mobile}. The best model, determined by the highest Coherence Score, was subject to a detailed review of its topic-word distributions. This phase involved a qualitative assessment of the terms most strongly associated with each topic, assigning labels that accurately reflected the aggregated meanings of these terms. To ensure focus on the most significant topics in further analyses, we applied a probability threshold, discarding any topics represented with less than 0.1 probability value in the document.

Consequently, the LDA model is trained by iterating through combinations of $(K, \alpha, \beta)$ and selecting the top five configurations with the highest coherence score for comparison. To validate the results, the topics were manually reviewed and the corresponding forum posts for the five candidate settings, and through careful verification, the optimal $K$ value was determined. Through these steps, we found that the tuple $(20,0.01,0.02)$ provided relatively detailed topics for the collected posts on quantum software programming from the Q\&A platforms. To assign appropriate descriptive labels to each discussion topic, the ten most frequent terms are particularly examined and, following the practices outlined in previous approaches \cite{ahmed2018concurrency, le2021large, zahedi2020mining}, reviewed the top 20 posts with the highest relevance to each topic.

\section{Results} \label{Sec:Results}
This section provides a complete summary of the data analysis results, which aim to comprehensively address the main RQs (research questions) of this study.

\subsection{RQ1: quantum software programming topics} \label{Sec: RQ1: Topics}

To answer RQ1, the topic modeling LDA approach (See Section \ref{Topic Modelling}) is used to analyse the data set and determine the optimal number of topics ($K=20$) and parameters ($\alpha=0.01$, $\beta=0.02$) by comparing coherence scores and conducting manual reviews of topic keywords and corresponding documents. After training the LDA model, the top 10 keywords are reviewed and analysed and read through the top 20 posts with the highest relevance to each topic. Based on the keywords identified by the LDA algorithm, related topics are manually identified.

\begin{figure}
    \centering
    \includegraphics[width=1.1\linewidth]{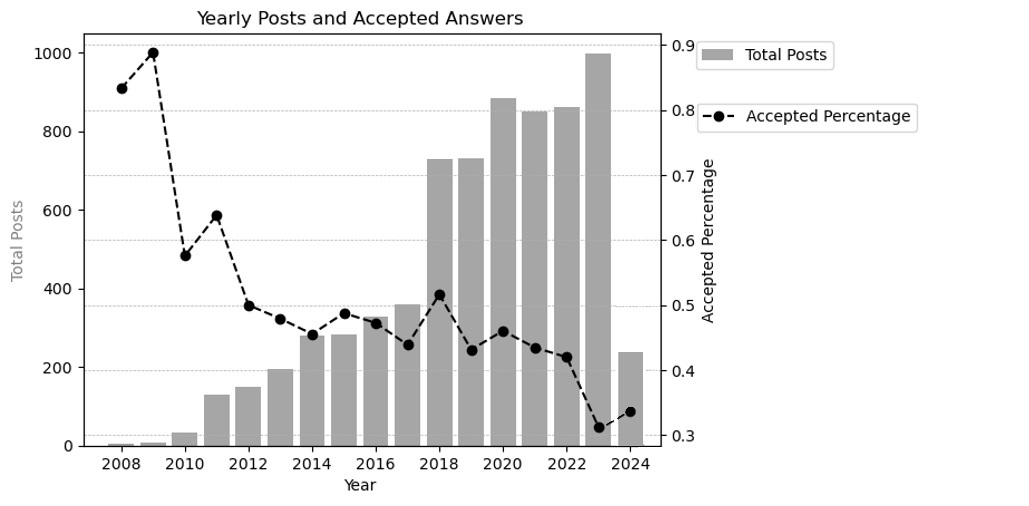}
    \caption{Total number of posts per year along with the proportion of questions whose answers were accepted}
    \label{fig:yearly-posts}
\end{figure}

After completing these steps, each topic is descriptively labeled, as shown in Table~\ref{tab:topics}. To filter less significant topics, we considered only those whose document-topic probability was at least 10\% as relevant \cite{barua2014developers}. Table~\ref{tab:topics} presents the complete set of identified topics, while Figure \ref{fig:yearly-posts} illustrates the total number of posts per year along with the proportion of questions whose answers were accepted. The bars represent the total number of posts made each year, while the line indicates the percentage of these questions that received accepted answers. The graph illustrates annual trends in the volume of questions posted and the community's responsiveness in terms of accepting answers. The growth in the total number of posts over time indicates an increasing interest and activity in quantum software programming. The proportion of accepted answers is used to assess the community engagement and efficacy in resolving issues. A decrease in the acceptable percentage suggests a rising complexity and difficulty in questions and a need for more expert participation. 

\begin{table}[h!]
\centering
\caption{Quantum Software Programming topics on Stack Exchange identified by LDA along with their proportions and trends over time.}
\label{tab:topics}
\begin{tabular}{|l|p{7cm}|p{6cm}|}
\hline
\textbf{Topic} & \textbf{Topic Name} & \textbf{Number (Proportion) of questions} \\ \hline
T0 & Python in Quantum Computing & 99 (1.43\%) \\ \hline
T1 & Quantum Devices and Finite Systems & 157 (2.27\%) \\ \hline
T2 & Quantum Code Management & 428 (6.18\%) \\ \hline
T3 & Parameter Control in Quantum Algorithms & 54 (0.78\%) \\ \hline
T4 & Object-Oriented Programming in Quantum Development & 117 (1.67\%) \\ \hline
T5 & Function and String Manipulation in Quantum Programming & 334 (4.82\%) \\ \hline
T6 & Quantum Circuit Simulation & 347 (5.01\%) \\ \hline
T7 & Quantum State Analysis and Measurement & 303 (4.38\%) \\ \hline
T8 & Security and Encryption Algorithms in Quantum Computing & 115 (1.66\%) \\ \hline
T9 & File Management and System Configuration in Quantum Computing & 407 (5.87\%) \\ \hline
T10 & General Quantum Computing Discussions & 1687 (24.32\%) \\ \hline
T11 & Package Management and Organization in Quantum Software & 76 (1.10\%) \\ \hline
T12 & Quantum Processing and Threading Management & 117 (1.69\%) \\ \hline
T13 & Quantum Data Processing and Logging & 112 (1.62\%) \\ \hline
T14 & Quantum States and Operations & 988 (14.24\%) \\ \hline
T15 & Documentation in Quantum Software Development & 80 (1.15\%) \\ \hline
T16 & Data Handling and Module Management & 677 (9.76\%) \\ \hline
T17 & Software Development Practices in Quantum Computing & 334 (4.81\%) \\ \hline
T18 & Physical Theories and Mathematical Foundations & 317 (4.58\%) \\ \hline
T19 & Programming Principles in Quantum Software & 196 (2.43\%) \\ \hline
\end{tabular}
\end{table}

Now, in the subsequent section, each identified topic (Table \ref{tab:topics}) is explained by providing sample developers' posts to enhance their understanding.

\begin{itemize}
    \item \textbf{Python in Quantum Computing (T0):} The code "T0" presents the developers' discussion related to the topic of the use of Python in the field of quantum computing. Python is a widely adopted programming language for developing quantum code and circuits. It provides several tools and frameworks that help with the implementation and evaluation of quantum algorithms. The practitioners asked questions about Python libraries such as Qiskit (\textit{How to get statevector() with defined gates in Qiskit? QC-PostID: \#16786}) and Cirq (\textit{In Cirq, how do you display circuit diagrams that are "prettier" than the ones displayed by default? QC-PostID: \#16521}), simplifying the process of implementing and evaluating quantum algorithms.

    \item \textbf{Quantum Devices and Finite Systems (T1):} Topic "T1" explores the practical aspects of quantum devices and systems with finite capabilities, focusing on hardware limitations, integration challenges, and system scalability in the context of quantum programming. Discussions include how constraints like gate speed, error rates, and coherence times affect quantum circuit performance and scalability. For example, questions like \textit{(What limits the speed of quantum gates? (QC-PostID: \#10057)}  emphasize the need to optimize quantum algorithms for real-world hardware, bridging the gap between theoretical models and practical implementation.

    \item \textbf{Quantum Code Management (T2):} Topic "T2" refers to the management and organisation of quantum codebases, with a focus on version control and hybrid methods adapted for quantum software development. For example, practitioners have discussed the existence of measurement standards and units for identifying qubits, such as (\textit{Are there measuring standards (and units) for the identification of qubits? QC-PostID: \#1889}).

    \item \textbf{Parameter Control in Quantum Algorithms (T3):} The topic "T3" focuses on the precise modification and optimization of parameters within quantum algorithms to enhance their performance and accuracy. Discussions cover techniques for parameter tuning and their impact on algorithmic outcomes, emphasizing practical applications in quantum programming. For instance, questions like \textit{(Error 400 on Variation Quantum Algorithm. QC–PostID: \#38672)} and \textit{(What exactly are "variational quantum algorithms"? QC–PostID: \#31640)} highlight challenges and solutions in implementing variational quantum algorithms, a core component of parameterized quantum circuits used in optimization and machine learning tasks.

    \item \textbf{Object-Oriented Programming in Quantum Development (T4):} This topic is related to the implementation of object-oriented programming techniques for developing quantum software. The process involves structuring quantum software projects to improve modularity and maintainability. Relevant post examples are (\textit{Qiskit: Can I run real experiments in QPUs with OOP structured code? QC–PostID: \#27867}).

    \item \textbf{Function and String Manipulation in Quantum Programming (T5):} The code "T5" relates to specific challenges and solutions in handling functions and strings in quantum software programming. It consists of problems related to encoding issues, the design of functions, and approaches for manipulation. Examples of posts include (\textit{How to add scalar to quantum string. QC–PostID: \#7216}) and (\textit{Compiling a classical function to a quantum circuit in practice. QC–PostID: \#12507}).

    \item \textbf{Quantum Circuit Simulation (T6):} The code "T6" focuses on discussions related to simulating quantum circuits and predicting the performance of quantum computers. It consists of quantum software applications and simulation approaches developed for the mentioned aim. For example, the developers asked (\textit{How exactly is solving the random circuit sampling problem a computation in the Church-Turing thesis sense? QC–PostID: \#8620}) and (\textit{Reversible Circuit Simulation (for classical tests of quantum subroutines) QC–PostID: \#6959}).

    \item \textbf{Quantum State Analysis and Measurement (T7):} The code "T7" is related to the techniques used for examining and characterizing the quantum states in quantum computing. Verifying the accuracy and viability of quantum computing has the highest priority. It verifies the performance of quantum algorithms and evaluates their viability for practical applications. The example post includes (\textit{Does Q\# maintain the state of a qubit, after performing the measurement? SO–PostID: \#54333024}).

    \item \textbf{Security and Encryption Algorithms in Quantum Computing (T8):} The "T8" topic encapsulates the developers discussions on developments and challenges faced by creating security and encryption algorithms for leveraging the full potential of quantum computing capabilities, including quantum key distribution (QKD) and post-quantum cryptography. For instance, developers asked questions such as (\textit{Is there a notion of "computational security" in quantum cryptography? QC–PostID: \#12812}) and (\textit{Doubt in Simple proof of Security of the BB84 QKD. QC–PostID: \#21629}).

    \item \textbf{File Management and System Configuration in Quantum Computing (T9):} The topic "T9" consists of developers posts related to the management of files and the configuration of systems designed particularly for quantum computing programming environments. This topic focuses on the unique system requirements and configuration challenges. Example post include: (\textit{How to use external .csv data file in quantum support vector machine Qiskit python code? QC–PostID: \#9967}).

    \item \textbf{General Quantum Computing Discussions (T10):} The code "T10" refers to a wide range of general and broad discussions related to quantum computing programming, which include fundamental concepts, emerging technologies, and challenges faced by the programming community. For example (\textit{Resources to get started programming quantum computers for non-physics majors. QC–ID: \#1367}) and (\textit{How can I save quantum gates as a graphic in png/svg format using Qiskit? SO–ID: \#78799576}).

    \item \textbf{Package Management and Organization in Quantum Software (T11):} The code "T11" deals with the organisation and management of packages in quantum software development, with a particular focus on optimizing dependency management and implementing a modular design. For example (\textit{Is it possible to turn modular multiplication into in-place operation? QC–PostID: \#26240}).

    \item \textbf{Quantum Processing and Threading Management (T12):} The "T12" topic addresses the management of computational tasks and optimization of threading resources effectively. For example, the developers discussed (\textit{How to reduce the length of Q\# programs execution time? QC–PostID: \#9506}).

    \item \textbf{Quantum Data Processing and Logging (T13):} The code "T13" consists of discussions that cover various aspects of data processing, encoding, and logging in quantum software systems, including how data is handled in a quantum context. For example (\textit{Encoding Numeric data as a quantum circuit in TFQ. QC–PostID: \#20707}).

    \item \textbf{Quantum States and Operations (T14):} The code "T14" explores the manipulation and control of quantum states and the execution of quantum operations, which are crucial for quantum programming. Discussion examples include (\textit{Increase execution speed of qiskit based python code for qasm\_simulator. QC–PostID: \#37321}) and (\textit{Can you find how many one qubit and 2 qubit gates are in a circuit in Qiskit? QC–PostID: \#27329}).

    \item \textbf{Documentation in Quantum Software Development (T15):} The code "T15" highlights the importance of documentation in developing and maintaining quantum software, addressing the best practices for documenting quantum code and architectures. An example post involves (\textit{The Bernstein Vazirani Algorithm widget (bv\_widget) in the Qiskit documentation. QC–PostID: \#17281}).

    \item \textbf{Data Handling and Module Management (T16):} The code "T16" specifically relates to managing data and modular components in quantum software programming projects. It focuses on exploring techniques for optimizing data flow and integrating components efficiently. Examples include posts such as (\textit{The queue\_info() module in Qiskit. QC–PostID: \#22091}) and (\textit{Cannot import tensorflow\_quantum module in Colab. QC–PostID: \#33994}).

    \item \textbf{Software Development Practices in Quantum Computing (T17):} The code "T17" consists of various posts related to software development approaches defined in the domain of quantum computing, including development practices and processes. For example (\textit{Conceptual: solving maze with multithreading. SO–PostID: \#17085043}) and (\textit{From an application programmer's perspective - Can Functional Programming be used to program Quantum Computers? QC–PostID: \#32037970}).

    \item \textbf{Physical Theories and Mathematical Foundations (T18):} The code "T18" encapsulates the fundamental principles of quantum computing, consisting of the theories and mathematical models that facilitate quantum computing technologies, as seen in the question (\textit{e.g., How to build a Hilbert space when do we already have concrete quantum logic gates available? QC–PostID: \#15066}) and (\textit{e.g., What is a Quantum-computer:'s qubit state like and How fast can it be modified? SO–PostID: \#58086285}).

    \item \textbf{Programming Principles in Quantum Software (T19):} The code "T19" explores programming principles and paradigms for developing quantum software. It specifically focuses on concepts like quantum logic gates and code entanglement. Example posts include (\textit{Error while compiling Quipper 0.7. SO–PostID: \#35076495}) and (\textit{From an application programmer's perspective- Can Functional Programming be used to program Quantum Computers? SO-PostID: \#32037970}).
\end{itemize}

In summary, the results from RQ1 cover a broad range of quantum software programming areas, from foundational aspects such as \textit{“Physical Theories and Mathematical Foundations (T18)”} to practical concerns like \textit{“Quantum Data Processing and Logging (T13)”}. These topics represent the diverse and interdisciplinary nature of quantum software programming, encompassing challenges in algorithm development, hardware integration, software optimization, and practical implementation. The breadth of topics reflects the complexity of quantum software programming as a field that merges quantum mechanics, computational theory, and modern programming paradigms.
The analysis further highlights the prominence of topics such as \textit{“General Quantum Computing Discussions (T10)”}, \textit{“Quantum States and Operations (T14)”}, and \textit{“Data Handling and Module Management (T16)”}, which are the most frequently discussed areas in quantum software programming. The frequent focus on these topics suggests their significance to the community, as they address both the foundational and practical aspects of quantum software development. For instance, general discussions (T10) provide an entry point for developers seeking guidance on diverse quantum programming challenges, while quantum states and operations (T14) are critical to understanding the mechanics and execution of quantum algorithms. Similarly, the emphasis on data handling and module management (T16) highlights the need for robust strategies to manage the unique data structures and modularity required in quantum programming projects. The prominence of these topics reflects the community efforts to address challenges that are central to the design, execution, and optimization of quantum software.

The growth in the number of questions posted annually, as illustrated in Figure ~\ref{fig:yearly-posts}, underscores a rising interest in quantum software programming. This trend indicates that the field is rapidly gaining traction, driven by advancements in quantum computing hardware, software frameworks, and a growing demand for quantum-ready applications. However, the decline in the proportion of accepted answers over time suggests that the questions being posed are becoming increasingly complex. This likely reflects the challenges developers face as they transition from foundational understanding to tackling advanced topics, such as concurrency management, optimization, and hybrid quantum-classical programming. The rising complexity of questions underscores the need for more expert contributions and better tools to support developers in resolving difficult quantum software programming issues.

These findings emphasize the evolving nature of quantum software programming as a specialized and challenging field. They highlight the importance of fostering collaboration among quantum software developers, researchers, and practitioners to address key challenges such as algorithmic optimization, software-hardware integration, and the development of scalable quantum applications. Additionally, the results suggest a need for targeted educational resources, better tooling, and community-driven solutions to lower barriers to entry and enhance the efficiency of problem-solving in quantum software programming.

\subsection{RQ1.1: Popular and Difficult Topics}

For a more in-depth analysis of the identified topics discussed in Section \ref{Sec: RQ1: Topics} (RQ1), now this section focuses on the most popular and difficult topics. This analysis aims to determine the popular and difficult topics practitioners discussed across the selected Q\&A repositories. The concepts of popular and difficult topics are discussed in Section \ref{Sec:Research Questions} (RQ1.1).

For identifying the popular and difficult topics, the min-max normalization technique has been used. This approach streamlines the values to a consistent scale, guaranteeing an equitable comparison across various topics. Additional information regarding the normalization process is elaborated in Section \ref{Sec:Research Questions} (RQ1.1). This classification will guide experts in focusing their efforts on the most critical issues. It will also enhance the understanding of complex concepts in quantum software programming by highlighting the areas that pose the greatest challenges. This analysis streamlines efforts and optimizes resource allocation to address specific issues and areas of interest in quantum software development.

\begin{figure}
    \centering
    \includegraphics[width=1\linewidth]{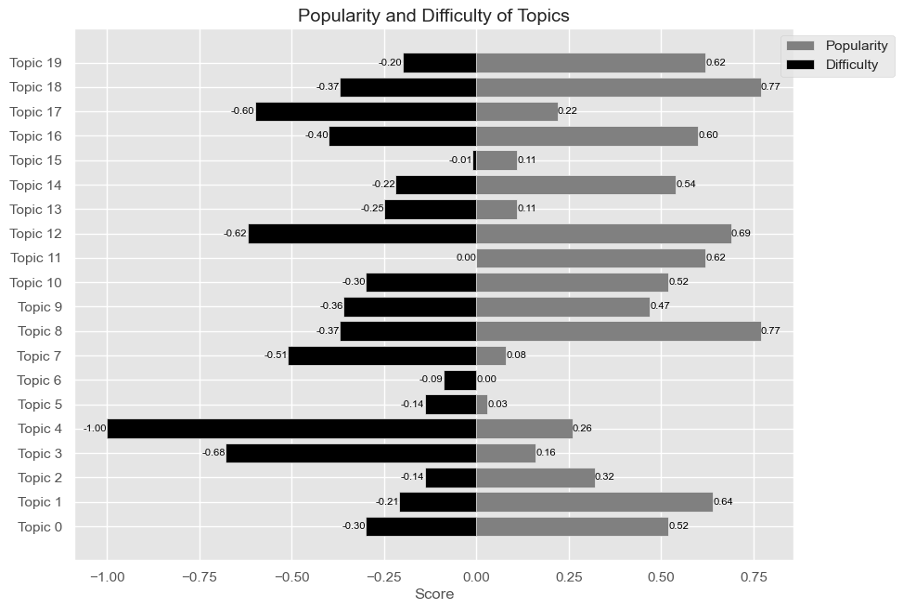}
    \caption{Popularity and difficulty of topics}
    \label{fig:pop-diff}
\end{figure}

This analysis leads us to identify the following popular and difficult topics:

\begin{itemize}
    \item \textbf{Topic popularity:} Using the min-max normalisation approach, the following three most popular topics on the selected Q\&A platforms are identified: \textit{“Security and Encryption Algorithms in Quantum Computing (T8)”, “Quantum Processing and Threading Management (T12)” and “Physical Theories and Mathematical Foundations (T18)”}, as illustrated in Figure \ref{fig:pop-diff}. Although T18 is the most popular topic, it is ranked eighth in terms of frequency distribution (see Table \ref{tab:topics}). Similarly, T8 is also the  most popular topic, and it is ranked thirteenth. The second most popular topic; T12 is ranked twelfth, as shown in Table \ref{tab:topics}. The Kendall’s Tau correlation test \cite{kendall1948rank} is applied to analyse the relationship between topic popularity and frequency distribution, and the resulting Tau value is 0.1319, indicating a weak positive correlation. The p-value acquired was 0.4171, which is significantly higher than the conventional threshold (typically 0.05) for establishing statistical significance.

    These statistics indicate that there is no significant association between the popularity of a topic and the frequency of its discussion across the Q\&A repositories. It reveals that the frequency of a topic does not correctly indicate its popularity—the total number of questions on a topic does not fully represent users' activities on Q\&A websites. Gaining an understanding of these findings allows software practitioners to better grasp the current trends in quantum software programming. 
    \item \textbf{Topic difficulty:} For topics difficulty, the following three specific topics of quantum software programming have been identified as difficult/complex for developers: \textit{“Object-Oriented Programming in Quantum Development (T4)”, “Parameter Control in Quantum Algorithms (T3)”, and “Quantum Processing and Threading Management (T12)”}. Similar to topic popularity, Kendall’s Tau correlation test is used to investigate the relationship between popularity and difficult topics. The analysis showed a tau value of -0.3430 with a p-value of 0.03, showing a significant negative association. This means that as the complexity of topics increases, their level of popularity on the Q\&A platforms declines. Specifically, less difficult topics tend to generate more attention and discussion, while more challenging topics, requiring a higher level of expertise to understand and engage with, tend to attract less discussion.
 \end{itemize}

Examining the popularity and difficulty of topics (RQ1.1) within quantum software programming has significant implications. The most popular topics in quantum software programming, such as \textit{“Physical Theories and Mathematical Foundations (T18)”, “Security and Encryption Algorithms in Quantum Computing (T8)” and “Quantum Processing and Threading Management (T12)”} reflect the growing interest in foundational and practical aspects of the field. These topics resonate with both new and experienced developers due to their broad applications and relevance to advancing quantum computing capabilities. Their popularity highlights the importance of basic quantum theories, information processing, ensuring robust encryption in a post-quantum world, and optimizing quantum software package management for better integration and performance. In contrast, the most difficult topics—\textit{“Object-Oriented Programming in Quantum Development (T4)”, “Parameter Control in Quantum Algorithms (T3)” and “Quantum Processing and Threading Management (T12)”}—underscore the steep learning curve in applying classical programming paradigms and managing quantum-specific complexities. Developers find it particularly challenging to define object-oriented practices, fine-tune algorithmic parameters, and handle concurrency in quantum environments. These difficulties emphasize the need for specialized tools and knowledge to advance the practical development of quantum software, as these areas are critical to scaling and improving performance in real-world applications.

Notably, \textit{“Quantum Processing and Threading Management (T12)”} stands out as both one of the most popular and one of the most difficult topics, highlighting a unique dual nature. On one hand, T12 popularity stems from its direct relevance to critical challenges in quantum computing, particularly its role in managing concurrency and optimizing quantum processes, which are important for real-world applications. On the other hand, its technical depth and the lack of tools contribute to its perceived difficulty. This duality suggests that while developers are keenly aware of its importance and engage with it out of necessity, they simultaneously struggle to master its complexities. Furthermore, the practitioners composition likely plays a role, with a mix of advanced and intermediate learners engaging with T12. Experts recognize its criticality in scaling quantum applications, whereas less experienced developers find it technically overwhelming. 

Moreover, the negative correlation observed between topics popularity and difficulty highlights the fact that although challenging topics are essential for the advanced understanding of quantum software programming, however, they are not usually discussed/addressed, likely due to the specific expertise required to contribute meaningfully to these discussions. This analysis reveals the fundamental inclination of developers to engage with content that is simpler to comprehend and implement, rather than struggling with more complex topics. It might be because of the fact that quantum programming is a relatively new and highly complex field. Most developers are still building foundational knowledge, so they tend to focus on simpler, more comprehensible topics, while avoiding more advanced areas that require specialized expertise and deeper understanding of quantum physics.

\subsection{RQ2: Quantum software programming tools and frameworks}
RQ2 is developed to explore the discussion posts within the selected dataset to identify the tools and framework used to support quantum software programming. The frequency of occurrences of each tool and framework was identified to understand how often they have been discussed. Tools and frameworks offer solutions-based platforms to support various quantum software development activities. 
Each identified tool and framework is evaluated based on the following five key criteria \cite{sajjad2018classification, khan2023software}, as listed in Table \ref{tab:quantum_tools}. \textbf{Source type} refers to whether a tool or framework is categorized as open source (OS) or closed source (CS). In open-source tools, the copyright holders grant users the rights to study, use, or modify the tool, framework, or system. \textbf{Input instructions} are the commands or parameters provided to execute the logic, categorized into three types: high-level (HL) instructions, quantum-specific instructions (QI), and mathematical variables (MV). The \textbf{output} refers to the results generated after execution, which can be classified as quantum source code (QSC), quantum algorithm (QA), or simulation findings (SF). \textbf{Automation level} indicates the degree of automation offered by the tool or framework, classified as fully automated (FA), semi-automated (SA), or non-automated (NA). Lastly, \textbf{evaluation} refers to the performance assessment of a tool or framework. It can be explicit (EX), where a comprehensive evaluation is performed, or implicit (IM), where only certain components are empirically assessed or evaluated to a limited extent.

\begin{table}[h!]
    \centering
    \caption{Quantum Software Programming Tools and Frameworks}
    \label{tab:quantum_tools}
    \resizebox{\textwidth}{!}{%
    \begin{tabular}{|l|l|l|l|l|l|l|}
        \hline
        \textbf{Tool/Framework} & \textbf{Source Type} & \textbf{Input Instructions} & \textbf{Output} & \textbf{Automation Level} & \textbf{Evaluation} \\ \hline
        MS QDK               & OS & HL       & QSC & FA & EX \\ \hline
        Ocean                & OS & QI       & SF  & FA & EX \\ \hline
        Strawberry Fields    & OS & HL       & SF  & FA & EX \\ \hline
        D-Wave Leap          & OS & QI       & SF  & SA & IM \\ \hline
        Cirq                 & OS & HL       & SF  & FA & EX \\ \hline
        Quantum Inspire      & OS & QI       & SF, QA & FA & EX \\ \hline
        PennyLane            & OS & HL       & SF  & FA & EX \\ \hline
        Forest               & OS & HL       & QSC & FA & EX \\ \hline
        Qiskit               & OS & HL, QI   & QA, SF & FA & EX \\ \hline
    \end{tabular}%
    }
\end{table}

\begin{figure}
    \centering
    \includegraphics[width=1\linewidth]{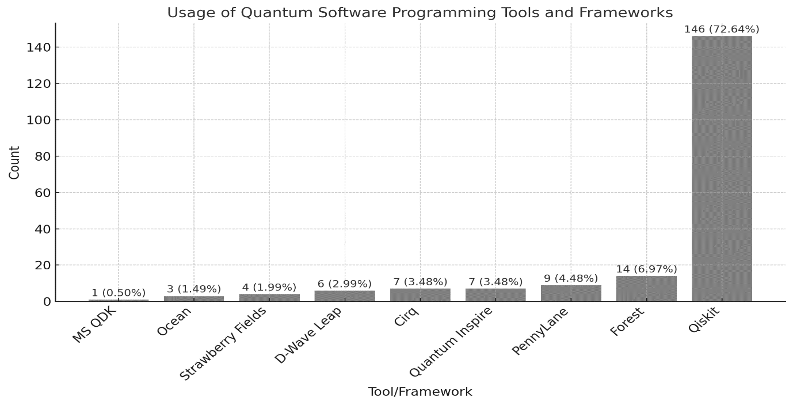}
    \caption{Distribution of tools and frameworks in quantum software programming}
    \label{fig:distribution_tools_framework}
\end{figure}

The results presented in Table \ref{tab:quantum_tools} reveal that all the identified tools and frameworks are open source. Similarly, six (n=6) identified tools and frameworks that execute input instructions or code written in a high-level (HL) programming format—these instructions are not explicitly dependent on the device executing them. Additionally, we found that seven (n=7) tools and frameworks generate output in the form of simulation findings (SF). Furthermore, eight (n=8) tools and frameworks are fully automated, and eight (n=8) have been explicitly evaluated. Moreover, we calculated the frequency and percentage of occurrence for each identified tool and framework in the selected dataset, as shown in Figure \ref{fig:distribution_tools_framework}. Based on the identified frequencies, we observed that \textit{Qiskit}\footnote{\url{https://www.ibm.com/quantum/qiskit}} is the most frequently discussed framework among practitioners across the selected Q\&A repositories, followed by \textit{Forest}\footnote{\url{https://github.com/rigetti/forest-software}} and \textit{PennyLane}\footnote{\url{https://pennylane.ai/}}. These three frameworks have been identified as the most frequently mentioned frameworks, highlighting their significance and extensive use within the quantum computing community.

Qiskit is an open-source quantum software development framework developed by IBM \cite{wille2019ibm}, designed for working with quantum computers at the level of extended quantum circuits, operators, and primitives. It allows users to perform a diverse set of quantum operations, including the development, execution, and simulation of quantum algorithms on both simulators and actual quantum hardware via IBM quantum devices \cite{wille2019ibm}. Qiskit consists of several components (Qiskit Terra, Aer, Ignis, Aqua) that cover the full-stack quantum software development process, from algorithm design to execution and simulation on quantum hardware. Built on Python, Qiskit makes it easy for classical developers familiar with Python programming to engage with quantum computing. Moreover, it includes built-in quantum algorithms specifically for machine learning, optimization, and quantum chemistry. Also, the Forest framework, developed by Rigetti Computing, provides a quantum software development and execution platform (including simulators and quantum devices) for quantum programs. Forest has four core components: Quil (Quantum Assembly Language), PyQuil (a Python-based high-level library that provides an interface to write and execute quantum code on Quil), QVM (Quantum Virtual Machine, which simulates quantum circuits), and QCS (Quantum Cloud Services, which offers Rigetti quantum devices as a cloud service). As a result, Forest is a full-stack, powerful hybrid-quantum computing platform that enables developers to write, run, and simulate quantum programs on quantum hardware \cite{olivares2020measurement}. Moreover, PennyLane is an open-source, Python-based framework developed by Xanadu that enables hybrid (classical-quantum) machine learning (ML). It seamlessly integrates with existing classical ML frameworks such as JAX, PyTorch, and TensorFlow, allowing developers to build hybrid ML models that can be executed on both quantum simulators and quantum devices. PennyLane primary feature is its ability to differentially program quantum circuits, meaning developers can use quantum circuits to train neural networks \cite{bergholm2018pennylane}. PennyLane provides a hardware-agnostic platform, enabling the execution of quantum circuits on various available quantum devices, such as processors from Xanadu, IBM, Google, and Amazon.

The findings of RQ2 provide significant insights into the existing quantum software development tools and frameworks. We observed that all the identified tools and frameworks are open source, and most of them use high-level programming formats that are device-agnostic. Most of these frameworks produce simulation-based output and are fully automated, with thorough evaluations. Additionally, we found that Qiskit, Forest, and PennyLane are the most frequently discussed frameworks across the selected Q\&A repositories. This highlights the widespread adoption of these three frameworks and their importance in the quantum software development ecosystem. Qiskit is particularly recognized for its extensive support of quantum software development, providing a full-stack infrastructure from algorithm design to execution, which makes it accessible to software developers. Forest offers a platform for hybrid (classical-quantum) computing by integrating both quantum simulators and devices. Similarly, PennyLane enables hybrid machine-learning workflows. These findings reveal the preferences of the quantum software development community and emphasize the most commonly adopted frameworks and tools for quantum software programming activities. The results of RQ2 help in making informed decisions when selecting appropriate tools and frameworks, ultimately supporting innovation and effective application in quantum software programming.

\subsection{RQ3: Quantum software programming challenges}
RQ3 aims to identify the challenges faced by practitioners in quantum software programming. We extensively analyze the data based on the research methodology discussed in Section \ref{Sec:Research Questions} (RQ3) and identify a list of 22 challenging factors. Using a thematic analysis approach (see Section \ref{Sec:Research Questions} (RQ3)), the identified challenges are further mapped across 9 sub-themes and 4 high-level themes (categories), as shown in Figure \ref{fig:mappings}. We now summarize the high-level themes, and the details are as follows:

\begin{figure}
    \centering
    \includegraphics[width=1.0\linewidth]{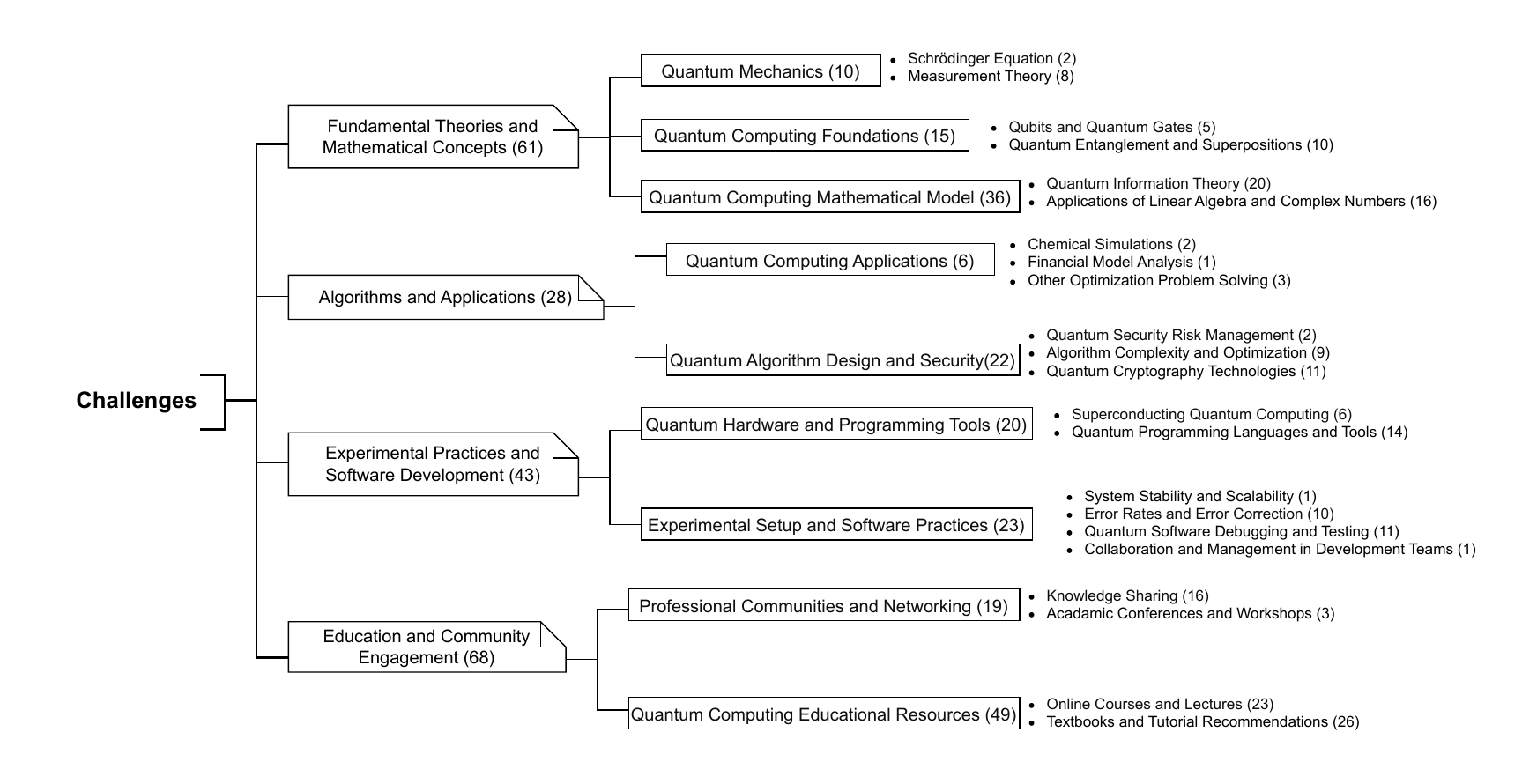}
    \caption{Thematic mapping of identified challenges}
    \label{fig:mappings}
\end{figure}

\subsubsection{Fundamental Theories and Mathematical Concepts}
A comprehensive analysis of challenges in quantum software programming revealed a strong focus on understanding the fundamental theories and mathematical principles of quantum mechanics and physics. By analyzing discussions on selected Q\&A sites, 61 posts among the top 200 specifically addressed theories and basic concepts of quantum mechanics. These discussions underscore the numerous questions and challenges practitioners face when grappling with the foundational concepts of quantum physics that underpin quantum computing. The analysis of these 61 posts identified six distinct challenges, categorized into three sub-themes (See Figure \ref{fig:mappings}). These challenges relate to the basic principles governing the operation of quantum systems, which are often counterintuitive and complex. This complexity makes it difficult for practitioners to fully understand and efficiently apply these principles within the context of quantum software development. For instance, quantum information theory, which governs how quantum systems process information, serves as the basis for designing and implementing quantum algorithms. Practitioners must translate abstract theoretical models, such as quantum superposition and entanglement, into practical constructs that can be programmed using quantum circuits and gates \cite{nielsen2010quantum}. Similarly, the representation and manipulation of quantum states in software often rely on advanced linear algebra techniques, such as eigenvector calculations and matrix operations, which are non-trivial for many developers without a strong mathematical background.
Additionally, understanding the Schrödinger Equation is crucial for simulating quantum systems, which is a key component in quantum software testing and debugging. However, its highly mathematical nature presents a steep learning curve \cite{berezin2012schrodinger}. Similarly, programming with qubits and quantum gates requires a deep understanding of their underlying mechanics, such as gate fidelity, coherence time, and error rates, which directly affect the reliability of quantum software \cite{preskill2018quantum}.

In summary, various challenges related to the fundamental concepts and theories of quantum mechanics reveal significant barriers that software practitioners face in the emerging field of quantum software programming. Practitioners must develop solid competencies in these fundamental quantum computing concepts, which require substantial training and educational resources. Progress in the broader adoption of quantum software development will be difficult without bridging the educational gap and creating more intuitive learning opportunities. Therefore, it becomes essential to provide comprehensive support systems for practitioners, including updating academic curricula and providing practical resources such as interactive simulators and programming toolkits to overcome these foundational challenges related to quantum computing knowledge.

\subsubsection{Algorithms and Applications}
In this category, the developers posts focusing on the quantum algorithms and their applications are analyzed. We noticed that 28 posts have preliminarily discussed the challenges related to quantum algorithms and their applications. We identified a total of six challenges which were further mapped into two sub-themes (See Figure \ref{fig:mappings}). 

The study findings revealed that \textit{quantum cryptography technologies} are considered a frequently discussed challenge by practitioners. These technologies present significant barriers in terms of implementation, specifically in key distribution and error correction techniques. Furthermore, the scalability of quantum cryptography systems and ensuring their compatibility with existing classical infrastructures persist as key barriers to extensive adoption \cite{scarani2009security}. We also identified that \textit{algorithm complexity and optimization} is highlighted by practitioners as the second most frequently discussed challenge. This is because quantum algorithms are inherently more complex than classical algorithms. Developing effective algorithms based on quantum information concepts poses a significant hurdle. Moreover, using these algorithms across specific quantum device platforms adds another layer of difficulty, as the limitations of existing noisy intermediate-scale quantum computers require careful tailoring of algorithmic designs to achieve practical results \cite{preskill2018quantum}.  Additionally, challenges such as \textit{chemical simulation} and \textit{other optimization problem solving} were identified at the application level of quantum algorithms. These present difficulties in precisely simulating complex molecular structures due to the limitations of existing quantum algorithms. Furthermore, optimization problem-solving in quantum computing remains an ongoing challenge, as finding optimal solutions for large-scale problems is constrained by both algorithmic efficiency and quantum device capabilities. Considering the complexity of quantum algorithms along with the limitations of existing quantum computers, substantial improvements are required to make these technologies robust and scalable. It necessitates strong interdisciplinary collaboration between practitioners from industry, computer systems engineers, and algorithm developers. Broadly, this category underscores the significance of continuous research and innovation to improve the effectiveness and applicability of quantum algorithms in real-world settings, eventually providing a roadmap to more robust quantum solutions across industries.

\subsubsection{Experimental Practices and Software Development}
This category is based on the challenges discussed by practitioners related to quantum software experimentation setups and development practices. We identified a total of 43 posts that explicitly focused on these challenges. Total six challenges are mapped in this category, with four of them classified under the \textit{experimental setup and software practices} sub-theme. This indicates that practitioners are particularly concerned with the experimental setting of quantum software programming and related practices. For example, \textit{quantum software debugging and testing} emerged as one of the most frequently discussed challenge in the mentioned sub-theme. Debugging and testing quantum software is inherently difficult due to the probabilistic nature of quantum computations, which produce non-deterministic outputs and make error tracing particularly challenging. Additionally, the limitations of existing noisy intermediate-scale quantum (NISQ) devices, such as restricted qubit availability and high noise levels, intensify these issues. Practitioners often rely on quantum simulators as substitutes for hardware access, but these simulators may not fully capture the behavior of real quantum systems, further complicating debugging efforts. The high error rates and lack of standardized error correction tools highlight the need for improved developer resources \cite{ramalho2024testing, preskill2018quantum}. In addition, the category touches upon \textit{system stability and scalability} and the challenges of \textit{collaboration and management in development teams }. Although these issues were less frequently discussed, they are critical for the broader adoption of quantum software development, particularly as the field moves toward more complex, large-scale applications.

Two challenges are classified across the \textit{quantum hardware and programming tools} sub-theme. As quantum software programming is intrinsically tied to the capabilities and constraints of quantum hardware. Quantum programming tools must optimize software for hardware-specific features, such as qubit connectivity, coherence times, and gate fidelities, which directly impact the performance of quantum programs. Furthermore, the high noise levels and instability of NISQ devices require tools that integrate hardware-aware features, including error correction protocols and resource-efficient execution strategies \cite{preskill2018quantum}. Simulators play a crucial role in bridging the gap between experimentation and real hardware usage, but their limitations in replicating actual quantum hardware behavior further emphasize the need for robust testing environments \cite{preskill2018quantum}.

Overall, we observed that \textit{quantum programming languages and tools} is the most frequently discussed challenge in this category, reflecting the practitioner community focus on finding more robust and operational tools to support the execution of quantum programs. These tools must not only provide efficient abstractions but also integrate capabilities for debugging, error correction, and optimization. Practitioners emphasized the need for more intuitive quantum programming tools that bridge the gap between classical and quantum development environments. 

In summary, the \textit{experimental practices and software development} category highlights a wide range of technical and operational challenges that directly impact the efficiency and effectiveness of quantum software programming. Addressing these challenges will require significant advancements in quantum programming tools, domain-specific programming languages, error correction techniques, and developer support. These improvements are essential for transitioning quantum programming from experimental research to scalable, real-world applications across industries.

\subsubsection{Education and Community Engagement}
In this category, the practitioners discussion posts related to the role of quantum education and community engagement were analyzed. We identified a total of 68 posts that significantly focused on both formal educational resources and community interactions in supporting the advancement of quantum programming knowledge and skills. A total of four challenges are mapped in this category, which are further classified into two sub-themes: \textit{quantum computing education resources} and \textit{professional communities and networking}.

One of the most frequently discussed challenges in this category is \textit{textbooks and tutorial recommendations}. Practitioners highlighted the pressing need for educational resources that cater to a wide audience, covering both foundational concepts and advanced topics in quantum software programming. Specifically, there is a demand for tutorials that offer hands-on experience with quantum programming tools such as Qiskit, Cirq, and pyQuil, enabling learners to bridge the gap between classical and quantum programming paradigms \cite{haghparast2024innovative}. These resources should focus on practical aspects, including algorithm implementation, debugging, and error mitigation, to equip developers with the skills needed to handle real-world quantum software challenges.

Similarly, \textit{knowledge sharing} emerged as an important challenge within the \textit{professional communities and networking} sub-theme. Practitioners emphasized the need for structured platforms to foster collaboration, share best practices, and facilitate peer learning in quantum software programming. They suggested leveraging professional forums, conferences, workshops, and seminars to disseminate knowledge on topics such as algorithm design, hardware-specific programming, and debugging strategies. This collaborative approach is particularly critical in the fast-evolving field of quantum software, where practitioners must stay updated on the latest advancements.

Overall, this category underscores the need for accessible educational resources and structured knowledge-sharing platforms tailored specifically to quantum software programming. The call for textbooks, hands-on tutorials, and learning tools highlights the importance of bridging the gap between classical and quantum programming while ensuring practical applicability \cite{angara2021teaching}. Additionally, fostering professional networking through events and platforms can enhance collaboration, enable peer-to-peer learning, and promote the dissemination of knowledge. Addressing these challenges will play a crucial role in developing quantum programming expertise and adapting to the rapid advancements in quantum software development.

The results of RQ3 reveal that the challenges identified in quantum software programming span across fundamental theories, algorithm development, experimental practices, and educational and community engagement. Practitioners struggle with understanding complex quantum mechanics concepts, such as superposition, entanglement, and the Schrödinger Equation, which are critical for algorithm design and software development. Tailoring quantum algorithms to NISQ hardware limitations further adds to the complexity, requiring interdisciplinary collaboration and innovation to improve scalability and real-world applicability. Debugging and testing quantum software remains inherently difficult due to probabilistic computations and high error rates, highlighting the need for better debugging tools, error correction techniques, and simulators. Additionally, the lack of accessible educational resources and structured knowledge-sharing platforms underscores the importance of hands-on tutorials, interactive learning tools, and professional networking opportunities to equip developers with practical skills. Addressing these challenges holistically is essential to advancing quantum software programming and transitioning from experimental research to scalable, real-world applications.

\section{General Implications}
We thoroughly discussed the specific implications of each research question in Section \ref{Sec:Results}. However, the general research and industrial implications derived from this study are as follows:

\subsection{Research Implications}
The study findings present several key implications for the research community, which are discussed below:

\textbf{Highlighting Knowledge Gaps in Quantum Programming:} The study shows a clear lack of understanding in key areas of quantum programming, such as quantum mechanics/physics and algorithm development. This suggests that researchers need to consider the existing knowledge gaps and focus on developing educational resources and research in these areas. The complexity of quantum programming languages and the difficulty in configuring quantum environments highlight the need for structuring these technologies to make them more accessible to a broader audience of researchers.

\textbf{Revealing the Need for Accessible Research Tools:} The challenges faced by practitioners highlight the need for tools that can support quantum programming in this emerging field of quantum software engineering. These tools can enable deeper investigation into quantum software development without requiring access to actual quantum devices. Researchers can use the study insights to develop more quantum-specific programming tools.

\textbf{Emphasis on Collaborative Efforts:} The study findings reveal that practitioners often face similar challenges in quantum programming, suggesting that more collaborative research across institutions can tackle these issues. By sharing knowledge and resources, researchers can mutually handle the complexities of quantum programming.

\textbf{Indicating Under-researched Areas:} The study highlights that various fundamental aspects of quantum software engineering, such as debugging and testing quantum programs, remain underexplored. This suggests a need for more research efforts to tackle these persistent challenges in quantum software engineering.

Overall, this study stresses the significance of tackling foundational gaps in the field, providing toolchain support, and encouraging collaboration among researchers. These efforts will drive progress in quantum software engineering generally and quantum programming specifically, leading to more effective development and problem-solving in this evolving area. 

\subsection{Industrial Implications}
We now discuss the important implications of the study results for the industries involved in quantum software development:

\textbf{Urgency for Workforce Development:} The study highlights the challenges practitioners face, implying that industries need to invest more in workforce development. Companies must ensure that their employees receive adequate training to keep up with the evolving quantum software development field, as this will enable them to handle the complexities associated with quantum software programming.

\textbf{Strategic Industry-Academia Partnerships:} Given the early stage of quantum software programming, industries can benefit from collaborating with academic institutions to bridge the knowledge gap. Such partnerships can ensure that both current employees and future professionals are well-prepared with the latest advancements in quantum software development.


\textbf{Open-Source Contributions:} The benefits of knowledge sharing in quantum software programming are clear. Industries should actively participate in and contribute to open-source projects, fostering innovation and exchanging ideas that can address common challenges, accelerate development, and maintain competitiveness in this emerging domain.

In summary, the study emphasize the need for workforce development, ensuring employees are trained to handle the complexities of quantum software programming. Strategic partnerships with academic institutions are crucial to bridging knowledge gaps. Additionally, active participation in open-source projects will foster innovation and help industries stay competitive in the rapidly evolving quantum software field.

\section{Threats to Validity}
The validity of this research could be influenced by several potential threats that might impact the reliability and generalizability of the findings. These threats are classified following the guidelines by Wohlin et al. \cite{wohlin2012experimentation}—internal validity, external validity, construct validity, and conclusion validity.

\subsection{Internal validity}
Internal validity refers to the degree to which certain factors influence the findings of the extracted data. In this empirical study, we identified various internal validity threats that appeared during different phases, as detailed below:


\textbf{Topic modeling:} We identified several internal validity threats related to the topic modeling approach. First, subjectivity in choosing the number of topics (\textit{K}) and tuning parameters like alpha (\(\alpha\)) and beta (\(\beta\)) could influence the results, as different values can lead to different topic structures. Furthermore, manually labelling the topics based on high-probability words involves judgment that may affect the accuracy and consistency of the identified themes. The preprocessing steps, such as removing low-frequency words and applying lemmatization and stemming, could result in missing the important context, potentially affecting the quality of the topics. To minimize these threats, coherence scores were used to evaluate the model, and a manual review confirmed that the most logical and interpretable topics were carefully chosen.

\textbf{Qualitative data analysis:} Various internal validity threats arise from the thematic approach followed in this study due to subjectivity in the coding process. The authors manually identified and categorized the themes, relying on their interpretations, which can introduce personal bias. Additionally, mapping initial codes into broader categories might affect consistency. However, the authors followed Braun and Clarke's \cite{braun2006using} guidelines and used the NVivo tool to ensure a structured process. 

\subsection{External validity}
External validity refers to the degree to which the study findings can be generalized to other scenarios, populations, settings, and different contexts of study. We noticed the following threats to the external validity of this study:

\textbf{Selected Q\&A Repositories:} The core threat to the external validity of the study emerged from the selection process of the chosen Q\&A repositories, which include Stack Exchange sites- Stack Overflow, Software Engineering, Code Review, and Quantum Computing. Based on it, we might have missed discussion posts on other repositories such as GitHub, which potentially limited the generalizability of the findings. To minimize this, we focused on widely used and trusted platforms in the developer community, ensuring a robust dataset. In the future, we plan to extend the study by
including other platforms, such as GitHub and Reddit, to capture more diverse perspectives in the quantum computing community.

\textbf{Sample size:} The sample size limitation for the qualitative analysis is a potential threat to the external validity of the study findings. While 6,935 posts were analyzed, only 200 highly ranked posts were used for the qualitative analysis, which may limit generalizability. However, this risk was minimized by using the AMS metric proposed by Bajaj et al. \cite{bajaj2014mining} to select developers' posts as possible representatives of practitioners' discussions on Q\&A platforms.

\subsection{Construct validity}
Construct validity evaluates the extent to which the measures used in a study accurately measure the specific constructs/concepts of interest. The following are major construct validity threats:

\textbf{Data coding and labelling:} The coding and labelling of themes in the qualitative data analysis is a potential threat to the study construct validity. The process is based on the researchers' subjective judgments, which can result in misclassification or missing key aspects of the challenges (RQ3). However, the mentioned threat is mitigated by using the  NVivo tools to perform the mapping process. Moreover, the data coding process is collaboratively discussed by all the involved authors, which helps to minimize the risk of inter-personal bias and ensure consensus (Section \ref{Sec:Research Questions}, RQ3).

\textbf{Identification of difficult and popular topics:} Another construct validity threat arises from how difficult and popular topics are identified in the study. Difficulty is concluded from discussion patterns, but this might not fully capture the actual complexity experienced by developers. To address these threats, a dual metric approach with the geometric mean and min-max normalization (Section \ref{Sec:Research Questions}, RQ1.1) was used to ensure consistent scaling of identified topics across difficult and popular categories. 

\subsection{Conclusion validity}
Conclusion validity refers to the degree to which the conclusions drawn from a study are credible and logically supported by the data. In this study, conclusion validity threats primarily stem from limitations in data analysis techniques. The adopted topic modelling approach provides structured insights; however, it might oversimplify complex discussions or miss less frequent but significant challenges, leading to incomplete conclusions. Furthermore, the subjective nature of qualitative data analysis (thematic mapping) can introduce biases that affect the accuracy and reliability of the study conclusions. However, peer discussions and multiple validation rounds were conducted to reduce subjectivity and improve the reliability of conclusions. Similarly, to minimize the topic modelling limitations, we applied coherence scores and conducted manual reviews to ensure the model captured both frequent and less common but important challenges.

\section{Conclusions and Future Work}
The rapid advancement of quantum computing has led to increased interest in quantum software programming. However, existing research primarily focuses on theoretical aspects, algorithmic development, and hardware improvements, with limited empirical studies examining real-world practitioner discussions and trends. This gap leaves a lack of understanding of how quantum software programming is practiced and the challenges it entails. In this study, we aim to address this gap by analyzing 6,935 discussion posts from multiple Stack Exchange sites (Quantum Computing, Stack Overflow, Software Engineering, and Code Review) to uncover key topics, widely used tools and frameworks, and challenges in quantum software programming. Our findings reveal a broad spectrum of topics, with \textit{ General Quantum Computing Discussions (T10), Quantum States and Operations (T14)}  and \textit{Data Handling and Module Management (T16)} emerging as the most frequently discussed topics. The study also highlights that  \textit{Security and Encryption Algorithms in Quantum Computing (T8), Quantum Processing and Threading Management (T12)} and \textit{ Physical Theories and Mathematical Foundations (T18)} are among the most popular topics, indicating  the growing interest in both fundamental principles and practical applications of the field. In contrast, \textit{Object-Oriented Programming in Quantum Development (T4), Parameter Control in Quantum Algorithms (T3)} and \textit{ Quantum Processing and Threading Management (T12)} are identified as the most difficult topics. It reveals that practitioners find it particularly challenging to establish object-oriented principles, optimize algorithmic parameters, and manage concurrency in quantum software programming.
Furthermore, we identified \textit{Qiskit, Forest} and \textit{PennyLane} as the most commonly used quantum software development frameworks, reflecting their importance in facilitating quantum programming tasks and the preferences of community. Our study also categorizes the challenges associated with quantum software programming into four key themes: \textit { Fundamental Theories and Mathematical Concepts, Algorithms and Applications, Experimental Practices and Software Development}, and \textit{ Education and Community Engagement}. Overall, we noticed that \textit{textbooks and tutorial recommendations, online courses and lectures, quantum information theory}, and \textit{ applications of linear algebra and complex numbers} are the most frequently discussed challenges classified across the mentioned categories. These highlight the critical need for enhanced educational resources, improved theoretical foundations, and better tooling to support practitioners in overcoming the complexities of quantum software programming.

In the future, we plan to expand empirical investigations by incorporating additional data sources, such as GitHub. Additionally, qualitative studies, including interviews and surveys with quantum software practitioners and researchers, will be conducted to provide deeper insights into the real-world application of quantum software programming principles. Bridging these gaps will help the quantum computing community develop more robust programming practices, accelerating the practical adoption of quantum technologies across industries.

\bibliographystyle{ACM-Reference-Format}
\bibliography{Main}

\end{document}